\documentclass[]{spie}  

\usepackage{amsmath,amsfonts,amssymb}
\usepackage{graphicx}
\usepackage{float}
\usepackage[colorlinks=true, allcolors=blue]{hyperref}
\usepackage{placeins}
\usepackage{longtable}
\usepackage{tabularx}
\usepackage{booktabs} 
\title{ Simulation package for solving dynamic diffraction problems in deformed crystals using  beam propagation method. 
Examples: Bragg, Laue geometry, asymmetric reflections, bend crystals, dislocations, crystals with arbitrary shapes and strain distributions and time dependent problems}

\author[a,b]{Jacek  Krzywinski}
\author[a]{Aliaksei Halavanau}
\affil[a]{SLAC National Accelerator Laboratory, 2575 Sand Hill Road , Menlo Park CA 94025, USA}
\affil[b]{National Centre for Nuclear Research, Otwock-Świerk, ul. A. Sołtana 7, Poland} 

\authorinfo{Further author information: (Send correspondence to J. Krzywinski)\\J. Krzywinski.: E-mail: krzywins@gmail.com \\  A. Halavanau: E-mail: aliaksei@slac.stanford.edu}

\begin{document} 
\maketitle

\begin{abstract}
We demonstrate the use of the Fast Fourier Transform Beam Propagation Method (FFT BPM) to simulate dynamic diffraction effects, including scattering from deformed crystals with arbitrary shapes in Bragg, Laue, and asymmetric geometries. The method’s straightforward algorithm, combined with FFT, enables fast computation and is easy to implement in Python. It successfully reproduces literature results for bent crystals, dislocations, and finite-shaped crystals simulated using the Takagi–Taupin equations. Python implementations for each case are provided in a public GitHub repository, with the code structured for parallel computing.

\end{abstract}

\keywords{dynamical diffraction; Fourier optics;
beam propagation method.}

\section{INTRODUCTION}
\label{sec:intro}  

Recently, we developed a new approach for solving time-dependent dynamic diffraction problems in distorted crystals using the Fast Fourier Transform Beam Propagation Method (FFT BPM). Our publication on this topic \cite{Krzywinski2022} focused on describing the theoretical approach and included several examples in the Bragg geometry. Here, we provide a more detailed description of the numerical implementation of our theory in Python. This implementation is optimized for parallel computing, and the Python code is available in a public GitHub repository \cite{fftbpm}.

We will also present examples, including simulations of scattering from deformed crystals with arbitrary shapes in Bragg, Laue, and asymmetric geometries. The examples involving bent crystals and crystals with dislocations illustrate how to account for non-uniform strain distribution inside the crystal when simulating dynamic diffraction using the FFT BPM method. We benchmarked our code by successfully reproducing results from the literature for bent crystals, dislocations, and finite-shaped crystals simulated using the Takagi–Taupin equations.

For continuity, we will repeat some sections of our previous paper to maintain the reading flow. 

\section{Theoretical approach}

\label{sec:Theo}  
The scalar Helmholtz equation for a constant angular  frequency $\omega$:
\begin{equation}
\label{eq:fov1}
		\left(\nabla^2+k(n(x,y,z))^2\right)E(x,y,z)=0
\end{equation}
 can be written in the following form\cite{Hadley92,Ersoy2007}:
 \begin{equation}
\label{eq:fov2}
	\frac{d\psi}{d\hat{z}}=i\left(\sqrt{1+{\nabla_\bot}^2-\delta\epsilon(\hat{x},\hat{y},\hat{z})}-1\right)\psi
\end{equation}
 where $\hat{x},\hat{y},\hat{z}= kx,ky,kz $, $k$ is the wave vector,${\nabla_\bot}^2$ is the Laplace operator taken with respect to the transverse coordinates $\hat{x}, \hat{y}$, $  \delta\epsilon(\hat{x},\hat{y},\hat{z})=n(\hat{x},\hat{y},\hat{z})^2-\bar{n}^2 $, $n(\hat{x},\hat{y},\hat{z})$ is the refractive index, $\bar{n}$ is its average value and  $ E(x,y,z)=\psi(x,y,z)e^{-ikz}$.
In the case of  beams with narrow angular spectrum this equation can be approximated by:
 \begin{equation}
\label{eq:fov3}
	\frac{d\psi}{d\hat{z}}=i\left(\sqrt{1+{\nabla_\bot}^2}-\frac{\delta\epsilon(\hat{x},\hat{y},\hat{z})}{2\sqrt{1-\left\langle k_x^2+k_y^2 \right\rangle}}-1\right)\psi
\end{equation}	

Here the operator $\sqrt{1+{\nabla_\bot}^2}$ in the denominator of the second term of the r.h.s  was approximated by $\sqrt{1-\left\langle k_x^2+k_y^2 \right\rangle}$ where $\left\langle k_x^2+k_y^2 \right\rangle$ 
is the  average angular spectrum and $k_x^2+k_y^2=F[{\nabla_\bot}^2]$ is the Fourier transform of  the ${\nabla_\bot}^2$ operator.
Eq. \eqref{eq:fov3} leads to the following FFT BPM equation:
 \begin{equation}
\label{eq:fov4}
	\psi(\hat{x},\hat{y},\hat{z}+\Delta{\hat{z}}) \approx F^{-1}\left\{F\left[\psi(\hat{x},\hat{y},\hat{z})\right]e^{i\Delta{\hat{z}}\sqrt{1-k_x^2-k_y^2}}\right\}e^{i\Delta{\hat{z}}\frac{\delta\epsilon(\hat{x},\hat{y},\hat{z})}{2 \cos{\alpha}}}
\end{equation}
where $F$ and $F^{-1}$ denote Fourier and inverse Fourier transforms, $\cos{\alpha}=\sqrt{1-\left\langle k_x^2+k_y^2 \right\rangle}$. The angle $\alpha$ corresponds to the grazing incidence angle with respect to the propagation direction.  
We have previously applied Eq. \eqref{eq:fov4} to take into account dynamic diffraction effects when simulating  interaction of x-ray beams with gratings and multilayer optics \cite{Gaudin12,ANDREJCZUK2015,Morgan2015,bajtAIP}. 

   In a large class of dynamical diffraction problems, the angular spectrum of the scattered x-ray beams consists of two narrow bands centered around the incident and the reflection angles. The meaningful information is contained within these bands. One can remove the fast oscillating component related to inter-planar spacing from the physical picture and derive FFT BPM equations for slowly varying envelopes of transmitted and reflected beams. For simplicity, we are considering the 2D ($x,z$) case, and the generalization to the 3D case is straightforward. The procedure is outlined below. First, one can write the scattered x-ray field as a sum of two components:   
  \begin{equation}
\label{eq:fov6}
    	\psi(\hat{z},\hat{x})=\psi(\hat{z},\hat{x})_{+} + \psi(\hat{z},\hat{x})_{-}.
\end{equation}
  Then the slowly varying envelopes are defined as follows:
  \begin{equation}
\label{eq:fov7}
    	\widetilde{\psi}(\hat{z},\hat{x})_{+} = \psi(\hat{z},\hat{x})_{+}e^{-i\frac{k_{d}}{2}\hat{x}},\ \  \widetilde{\psi}(\hat{z},\hat{x})_{-} = \psi(\hat{z},\hat{x})_{-}e^{i\frac{k_{d}}{2}\hat{x}}
\end{equation}
where $k_d$ is the reciprocal vector related to inter-planar spacing.
Next we expand the exponential term related to dielectric susceptibility in Eq. \eqref{eq:fov4} as:
  \begin{equation}
\label{eq:fov9}
    e^{i\hat{z}\frac{\delta\epsilon(\hat{z},\hat{x})}{2cos{\alpha}}} = \sum_{n=-\infty}^{+\infty}\Delta\epsilon(\hat{z},\hat{x})_{n}e^{ik_{d}n\hat{x}}
\end{equation}
When crystal planes, taking part in the reflection, are perpendicular to the $x$-axis, one can describe the $\delta\epsilon(\hat{z},\hat{x})$ in Eq. \eqref{eq:fov9} as a complex harmonic function with the period equal to the inter-planar  spacing:
  \begin{equation}
\label{eq:fov8}
    \delta\epsilon(\hat{z},\hat{x}) = \chi_{0}+2\chi_h\cos{k_d\hat{x}}
\end{equation}
and $\chi_{0}=\chi_{0r}+i\chi_{0h}$ and $\chi_h=\lvert\chi_{rh}\rvert-i\lvert\chi_{ih}\rvert$.
The complex amplitude is composed of the $\chi_{0r}$, $\chi_{0h}$, $\chi_{rh}$ and $\chi_{ih}$ which 
could be found, for example, at the web page “x-ray dynamical diffraction data on the web” (https://x-server.gmca.aps.anl.gov/x0h.html) or XOP  x-ray optics software toolkit \cite{XOPcode}.

By applying the Jacobi–Anger expansion: $e^{i w \cos{\theta}} = \sum_{n=-\infty}^{\infty} i^n J_n(w) e^{i n \theta}$  the first three expansion terms in Eq. \eqref{eq:fov9} can be written in the following form: 
  \begin{equation}
\label{eq:fov11}
	\Delta\epsilon_{0}=J_{0}\left( z\frac{\chi_h}{\cos{\alpha}}\Pi(x,z)\right)e^{\frac{iz}{2\cos{\alpha}}\chi_{0}\Pi(x,z)}
\end{equation}

  \begin{equation}
\label{eq:fov12}
	\Delta\epsilon_{1}=\Delta\epsilon_{-1}=iJ_{1}\left(z\frac{\chi_h}{\cos{\alpha}}\Pi(x,z)\right)e^{\frac{iz}{2\cos{\alpha}}\chi_{0}\Pi(x,z)}
\end{equation}
where $\Pi(x,z) = 1$ inside the crystal,  and $\Pi(x,z) = 0$ outside the crystal, with $J_n(w)$ being Bessel functions. 

 We treat deformation of the crystal as modification of the susceptibility \cite{Authier2003-ae}:
 
   \begin{equation}
\label{eq:Authier}
    	\Delta\epsilon^{'}_{n}(\hat{z},\hat{x}) = \Delta\epsilon_{n}(\hat{z},\hat{x})e^{in\mathbf{k_{d}u}(\hat{z},\hat{x})}
\end{equation}
 where $\mathbf{k_{d}u}$ is a scalar product of the reciprocal  and displacement vectors. We note that  $\mathbf{k_{d}}$ is parallel to $\hat{x}$ by definition, therefore only $x$-component of $\mathbf{k_{d}}$ is used.

It has been shown in our previous paper \cite{Krzywinski2022} 
 that one arrives at the following set of equations for slowly varying envelopes $\widetilde{\psi}(\hat{z},\hat{x})_{+}$ and $\widetilde{\psi}(\hat{z},\hat{x})_{-}$: 

  \begin{equation}
  \begin{aligned}
\label{eq:fov10}
    \widetilde{\psi}(\hat{z}+\Delta\hat{z},\hat{x})_{+} =  \widetilde{\psi}(\hat{z},\hat{x})_{p+}\Delta\epsilon_0(\hat{z},\hat{x})+\widetilde{\psi}(\hat{z},\hat{x})_{p-}\Delta\epsilon^{'}_{+1}(\hat{z},\hat{x}) \\ 
        \widetilde{\psi}(\hat{z}+\Delta\hat{z},\hat{x})_{-} =  \widetilde{\psi}(\hat{z},\hat{x})_{p-}\Delta\epsilon_0(\hat{z},\hat{x})+\widetilde{\psi}(\hat{z},\hat{x})_{p+}\Delta\epsilon^{'}_{-1}(\hat{z},\hat{x}),
    \end{aligned}
\end{equation}
where $\widetilde{\psi}(\hat{z},\hat{x})_{p\pm}=F^{-1}\left[F\left[\widetilde{\psi}(\hat{z},\hat{x})_{\pm}\right]p_{\mp}\right]$.
The operators $p_{\pm}=e^{i\Delta{\hat{z}}\sqrt{1-(k_{x} \pm \frac{k_d}{2})^2}}$ correspond to the Fourier image of ${\nabla_\bot}^2$ which is shifted by $\pm \frac{k_d}{2}$ in the angular spectrum space.
 Equation \eqref{eq:fov4} can be rewritten in the operator form as: 
\begin{eqnarray}
\label{split-op}
    \psi(\hat{x},\hat{y},\hat{z}+\Delta{\hat{z}}) &\approx& \prod_i A(a_i \Delta_z) B (b_i \Delta_z) \psi(\hat{x},\hat{y},\hat{z}), \\ \nonumber
    A &=& F^{-1}\left\{F\left[\psi(\hat{x},\hat{y},\hat{z})\right]e^{i a_i\Delta{\hat{z}}\sqrt{1-k_x^2-k_y^2}}\right\},\\ \nonumber
    B &=& e^{i b_i\Delta{\hat{z}}\frac{\delta\epsilon(\hat{x},\hat{y},\hat{z})}{2 \cos{\alpha}}}.
\end{eqnarray}
In the case of first order splitting, when $a_1 = 1, b_1 = 1$, Eq. \eqref{split-op} is reduced to Eq. \eqref{eq:fov4}. For most of the practical problems this splitting scheme is accurate. Higher order splitting improves the efficiency of the FFT BPM for large Bragg angles. For instance, 
when $a_1 = 0.0, a_2 = 1.0, b_1 = 0.5, b_2 = 0.5$, Eq. \eqref{split-op} is accurate up to $O(\Delta z^2)$, and when $t = 1.3512, a_1 = 0.0, a_2 = t, a_3 = 1 - 2t, a_4 = t, b_1 = t/2, b_2 = (1 - t)/2, b_3 = (1 - t)/2, b_4 = t/2$ it is accurate up to $O(\Delta z^4)$; see \cite{OMELYAN2002188}.

.

We point out that Eqs. \eqref{eq:fov10} can be treated as a two-beam approximation for the FFT BPM. These equations are analogous to the Takagi-Taupin equations (TTE) in two-beam approximation \cite{Authier2003-ae}. However, there is a significant difference between TTE and FFT BPM equations. The TTE equations are a system of hyperbolic equations where the second derivatives in the transverse direction with respect to beam propagation are neglected. Therefore, diffraction of the x-rays is not taken into account in the TTE formulation. The numerical algorithms for solving TTE, which are presented in the literature, typically require setting the boundary conditions. This could be a difficult problem itself in complicated geometries or when the boundaries are not very well defined. On the other hand, the FFT BPM equations are a system of parabolic equations that automatically includes diffraction. Also, as we have mentioned before, the FFT BPM method is especially convenient when dealing with complicated shapes or non-homogeneous boundaries. 
\section{Numerical implementation}
The algorithm is embedded in a class; the user provides configuration file (in yaml format) to run a simulation instance, for example for a given angle of incidence and/or wavelength. Since most of the diffraction problems are "linear", all instances, in principle, can be evaluated in parallel. 
The input configuration file provides the main input  simulation parameters and data: photon energy, crystallographic data, electric susceptibilities, crystal geometry, numerical grid parameters, integration methods and parameters, x-ray beam properties, crystal deformation data and program settings. 

The crystal boundaries can be defined directly in the configuration file for simple cases, such as those in Bragg geometry, or by generating a mask with additional Jupyter notebooks and saving it to a file. We have provided examples of the latter approach for simulations of asymmetric reflections and scattering from 3D finite crystals with cuboid, cubic, and hemispherical shapes.    

The user interacts with the code using jupyter notebooks or user scripts (similar to SRW approach).  Python scripts are grouped in three classes \cite{fftbpm}: {\sc XBPM}, {\sc XCrystal} and {\sc XCrTools} 

The main propagation equations, Eqs. (9-14) from our previous paper \cite{Krzywinski2022}, are encoded in the {\sc XBPM} class. The {\sc XCrTools} class is used to configure the incident x-ray beam and to provide numerical tools needed for manipulations such as FFT, padding, cropping, interpolation, etc. The {\sc XCrystal} class incorporates the parameters and data provided in the yaml configuration file and creates the main 'run3D()' function. The 'run3D()' can be used to define a single realization reflection object that is computed in parallel to other reflections having different parameters e.g. photon energy, incidence angle, etc. Thus, the calculation of rocking curves  or time-dependent diffraction can be implemented as an embarrassingly parallel problem.

\section{Simulation of reflection in the Laue geometry and asymmetric reflections }

Previously we have shown simulations of the symmetric Bragg reflection cases. Now let us consider the Laue geometry and the diamond crystal for (400) reflection at 9.831 keV photon energy. The crystal has a form of a slab. Let us define the $s$-axis to be perpendicular to the crystal's surfaces.  The orientation of the $s$-axis with respect to the crystallographic planes is defined by the $asymm\_angle$ parameter in the configuration yaml file. The Bragg geometry corresponds to $asymm\_angle=0$ $deg$ and the Laue geometry corresponds to the $asymm\_angle=90$ $deg$.

\begin{figure}[H]
   \begin{center}
   \includegraphics[width=0.3\linewidth]{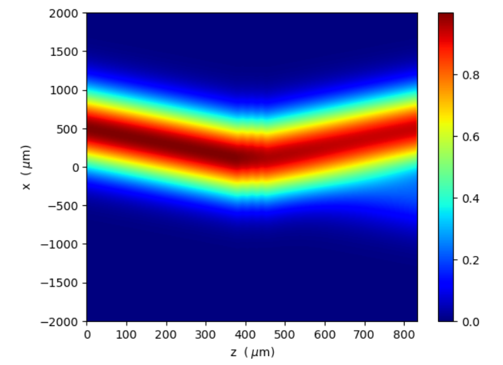}
   \includegraphics[width=0.3\linewidth]{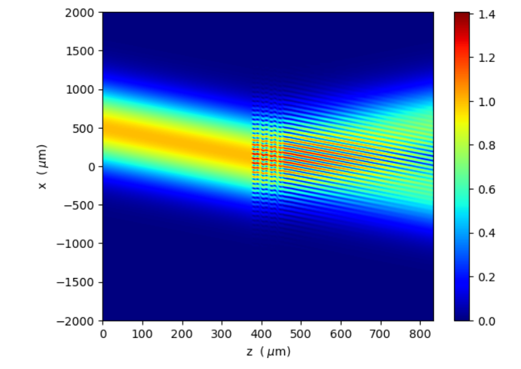}
   \\
   \includegraphics[width=0.3\linewidth]{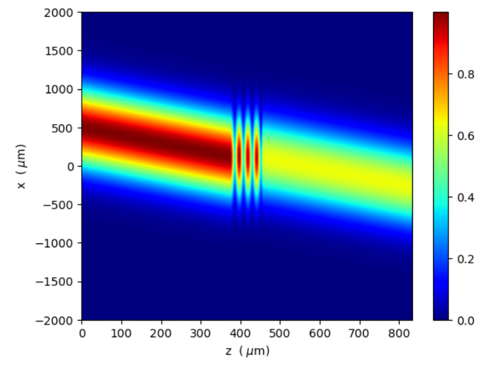}
   \includegraphics[width=0.3\linewidth]{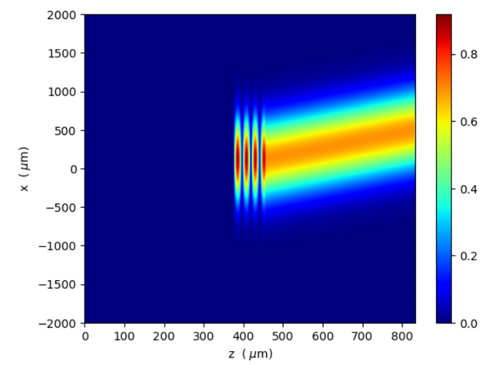}

   \end{center}
 
   \caption{2D visualization of a diamond (400) Laue reflection, with the incident beam at the Bragg angle (top left) and at the Bragg angle + 1.5 $\mu$rad (top right). The bottom row shows the intensity of the transmitted component $\widetilde{\psi}{_+}$ (left) and the reflected part $\widetilde{\psi}{_-}$ for the 1.5 $\mu$rad deviation from the Bragg angle. The Pendellösung oscillations are clearly visible. The color scale units are arbitrary}
   \label{fig:2Dvis}
   \end{figure}

An example of a simulation of reflection in the Laue geometry is implemented in a Jupyter notebook, which can be found in the \texttt{example} subdirectory of the Crystal-fft-bpm GitHub repository \cite{fftbpm}. The corresponding configuration file is located in the \texttt{config} subdirectory. The results of the simulation are presented in Fig. \ref{fig:2Dvis}. Two cases are simulated: when the incident beam is at the Bragg angle (top left) and at the Bragg angle + 1.5~$\mu$rad (top right).

The names of the Jupyter notebook and the configuration files are:
\begin{itemize}
    \item \texttt{SingleRealization\_C400\_Laue.ipynb}
    \item \texttt{C400\_9p8keV\_LaueFig1.yaml}
\end{itemize}

\begin{figure}[H]
   \begin{center}
   \includegraphics[width=0.3\linewidth,height=3.75cm, width=5.55cm]{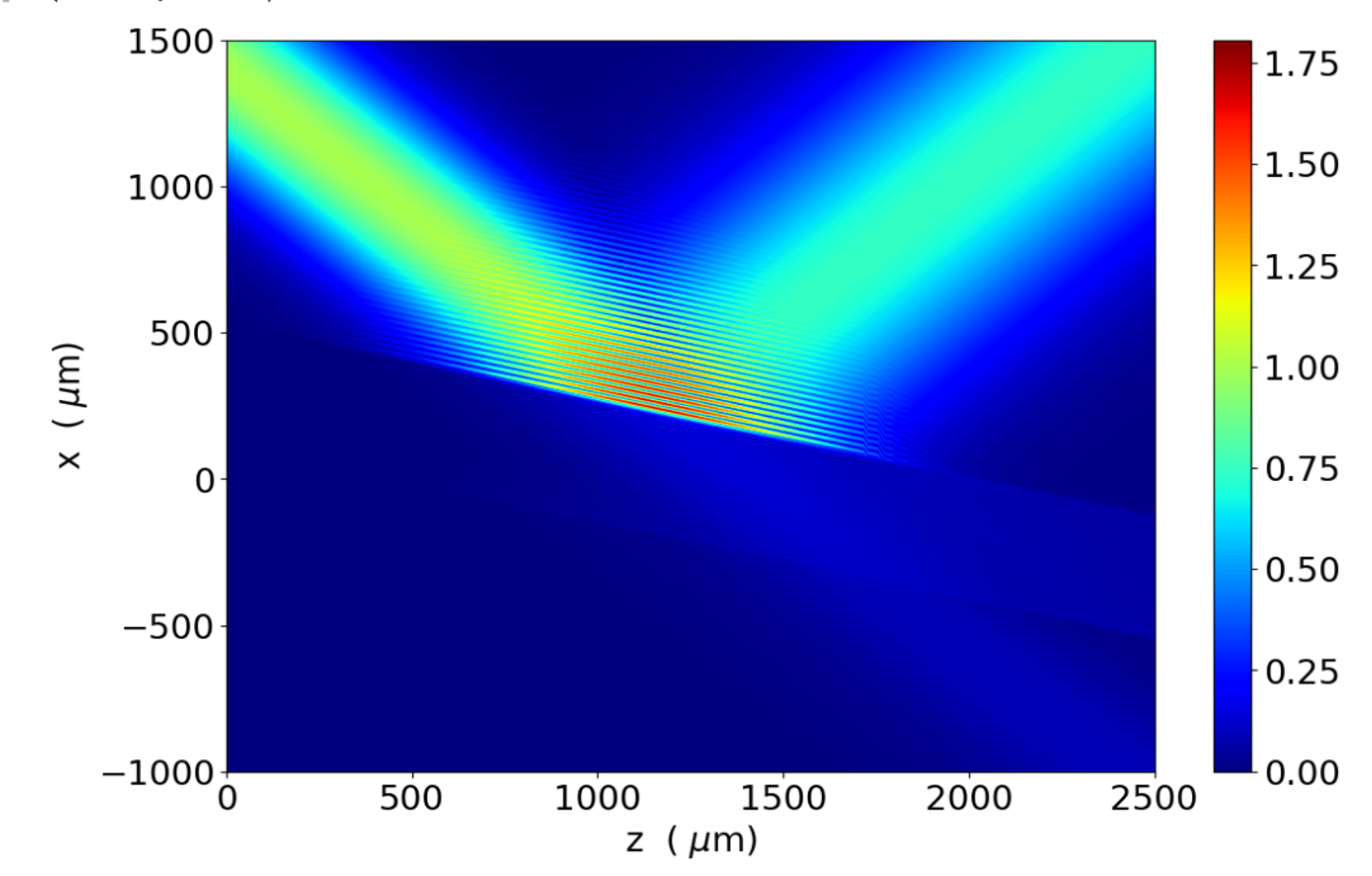}
      \includegraphics[width=0.3\linewidth,height=3.7cm, width=5.55cm]{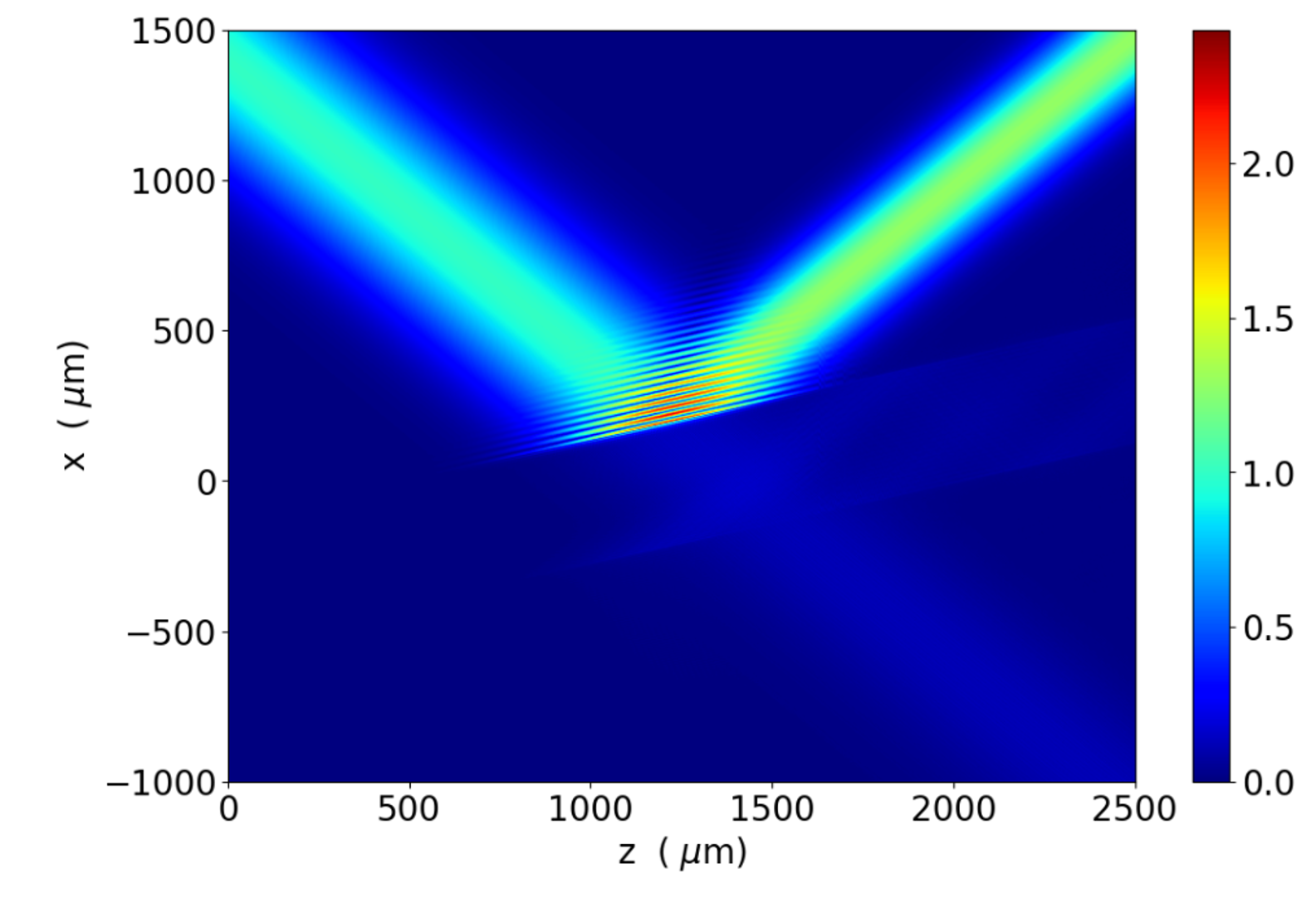}
   \end{center}
 
   \caption{2D visualization of field amplitudes for a diamond  (400) asymmetric reflection, asymm\_angle=15 $deg$ (left), asymm\_angle=15 $deg$ (right).  Color scale units are arbitrary.}
   \label{fig:2DvisAsymm}
   \end{figure}

Asymmetric reflection geometries occur when \(0^\circ < \texttt{asymm\_angle} < 90^\circ\). 
A visualization of asymmetric reflection for asymmetry angles of \(\texttt{asymm\_angle} = 15^\circ\) and \(-15^\circ\) is shown in Fig. \ref{fig:2DvisAsymm}. 
The deviations from the Bragg condition for these angles were 20 \(\mu\)rad and 10 \(\mu\)rad, respectively. 
The corresponding Jupyter notebooks for generating the crystal boundary mask, performing the beam propagation, and the configuration file for \(\texttt{asymm\_angle} = 15^\circ\) are:

\begin{itemize}
    \item \texttt{geometry-AsymetricReflection15deg.ipynb}
    \item \texttt{SingleRealization\_C400\_AsymetricReflection.ipynb}
    \item \texttt{C400\_9p8keV\_LaueAssymRefl15degFig2.yaml}
\end{itemize}

To run the simulation, the crystal boundary mask file must first be generated by executing the \\ \texttt{geometry-AsymetricReflection15deg.ipynb} notebook.

\begin{figure} [ht]
   \begin{center}
   \begin{tabular}{c} 
   \includegraphics[height=5cm]{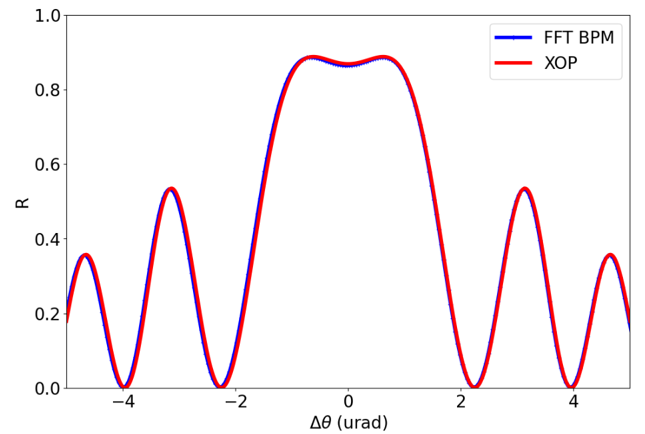}
   \end{tabular}
   \end{center}
   \caption[LaueRocking] 
   { \label{fig:LaueRocking} 
Rocking curve simulated in the Laue geometry for a 82 $\mu$m thick diamond crystal at 9.813 keV photon energy.}
   \end{figure}

The rocking curve for the Laue case was computed using the parallel algorithm outlined in the previous paragraph, with parallelization performed with respect to the incident angle. The results obtained using the FFT BPM method are compared to those of the XOP toolkit \cite{XOPcode}, as shown in Fig. \ref{fig:LaueRocking}. We note excellent agreement between our method and the XOP output. The very small differences between the two plots can be attributed to the finite beam waist and numerical errors due to the finite mesh size. The corresponding Python code for parallel computing, the Jupyter notebook for displaying the results, and the configuration file are: 
\begin{itemize}
 \item \texttt{run\_parallel\_angleC400\_9831eV\_Laue.py}
 \item \texttt{process-parallel-data-angleC400\_9p8keV\_Laue.ipynb}
 \item \texttt{C400\_9p8keV\_LaueFig3.yaml}.
\end{itemize}

   \FloatBarrier

\section{Simulations of Bend crystals }

Bend crystals have been successfully used to build single-shot spectrometers for XFEL applications. \cite{Feng, 10.1063/5.0019935}
To benchmark our code, we simulated rocking curves for a bend diamond crystal applied at the European X-ray Free-Electron Laser. The rocking curves of this crystal were theoretically investigated in the article by Samoylova et al.\cite{Samoylova:xn5016}. In that paper the simulations were done for a 20 $\mu m$ thick crystal bent around the axis [001] (y axis) with a surface orientation [110] (x axis). The displacement $u_x$ for the cylindrically bent crystal can be expressed as:

 \begin{equation}
\label{eq:fov110}
	u_x=\frac{1}{2R}\left(z^2+ \nu x^2\right)
\end{equation}
where R is the bending radius and $\nu$ is the Poisson ratio \cite{Samoylova:xn5016}.  The  scattering from the bend crystal was derived  by a numerical solution of the Takagi–Taupin equations based on the algorithm proposed by Authier et al \cite{Authier2003-ae}.
The authors considered a symmetric Bragg reflection 440 of monochromatic x-ray at 14.4 keV.  The incident wave front was restricted to a width of 10 $\mu m$ by an entrance slits. 

\begin{figure} [ht]
   \begin{center}
   \begin{tabular}{c} 
   
      \includegraphics[width=0.4\linewidth]{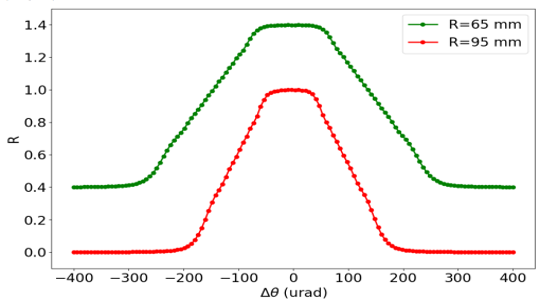}
      
   \end{tabular}
   \end{center}
   \caption[BendRocking] 
   { \label{fig:BendRocking} 
Rocking curves simulated for bending radii of 65 and 95 mm.}
   \end{figure}

We have simulated the same case using our FFTBPM method. The rocking curves for two bending radii of R= 65 mm and R= 95 mm are presented in Fig. \ref{fig:BendRocking} 

The results calculated using our FFT BPM method, presented in Fig.~\ref{fig:BendRocking}, are identical to the numerical solution of the Takagi–Taupin equations shown in Fig.~2 in the article by L. Samoylova et al.~\cite{Samoylova:xn5016}. The corresponding Python code for parallel computing, the Jupyter notebook for displaying the results, and the configuration files for $R = 65$ mm and $R = 95$ mm are:  

\begin{itemize}
  \item \texttt{run\_parallel\_angleC440CurvedSpectrometerR65.py}
   \item \texttt{run\_parallel\_angleC440CurvedSpectrometerR95.py}
  \item \texttt{process-parallel-data-angleC440Spectrometer.ipynb}
  \item \texttt{CrystalC440Fig4\_R65.yaml}
  \item \texttt{CrystalC440Fig4\_R95.yaml}
\end{itemize}

In order to run the notebook and display the results, one needs to run the Python code twice using either of the configuration files.

\section{SIMULATIONS OF DISLOCATIONS}
 Our FFT BPM method can be easily applied to simulate crystal defects such as dislocations. We will simulate two cases that were investigated using the TTE approach: a Laue [220] and [440] reflections from a Si crystal in the presence of screw and mixed dislocations. \cite{Authier2003-ae} \cite{Besedin2014-bk}, \cite{KowalskiGronkowski}. 
 
 The first case follows an example presented in the work of Besedin et al.\cite{Besedin2014-bk}. A plane-parallel silicon plate was chosen as the model crystal. In this structure, undissociated straight line dislocations have Burgers vectors with magnitude $\lvert \textbf{b} \rvert = a/\sqrt{2}$, where $a$ is the lattice constant. The Burgers vectors and the diffraction vector $\boldsymbol{h}$ are directed along the (110) axis. 

\begin{figure} [ht]
   \begin{center}
   \begin{tabular}{c} 
   \includegraphics[height=5cm]{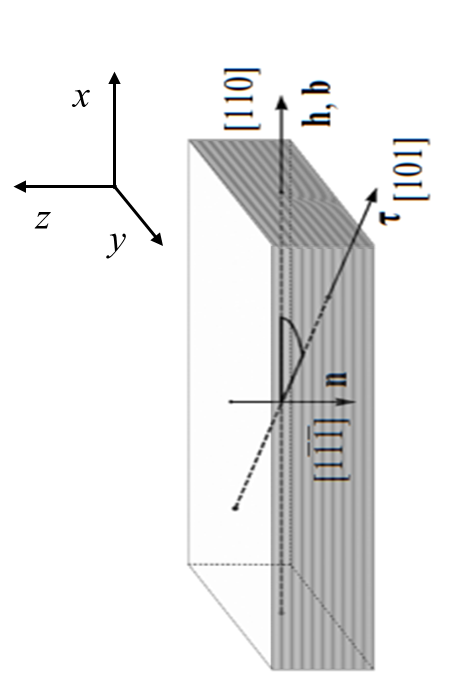}
   \end{tabular}
   \end{center}
   \caption[ConfigChuch] 
   { \label{fig:ConfigChuch} 
Geometry of the Laue [220]reflection from a Si crystal in the presence of screw and mixed dislocations. The screw dislocation has the unit dislocation line vector  $\boldsymbol{\tau}$  parallel to the Burgers vector (along the(110) axis). The  mixed 60 degrees dislocation has  the unit dislocation line vector  $\boldsymbol{\tau}$  parallel to the (101) axis. The pattern on the crystal sides indicates the Pendellösung 
fringes.}
   \end{figure}

We will consider two types of dislocations here: a screw dislocation and a mixed 60-degree dislocation. The screw dislocation has the unit dislocation line vector $\boldsymbol{\tau}$ parallel to the Burgers vector (along the (110) axis). The mixed 60-degree dislocation has the unit dislocation line vector $\boldsymbol{\tau}$ parallel to the (101) axis, as shown in Fig. \ref{fig:ConfigChuch}.

  According to \cite{Besedin2014-bk} the displacement vector u(r) for a straight-line dislocation in its intrinsic coordinate system ($\textit{$x_0$,$y_0$,$z_0$}$),
   has the form:
   \begin{equation}
\label{eq:fov120}
	\boldsymbol{u}= \frac{\boldsymbol{b}}{2\pi}\arctan\frac{z_0}{y_0}+\frac{\boldsymbol{b-\tau(b\cdot\tau})}{2\pi}\frac{y_0z_0}{2(1-\nu)(y_0^2+z_0^2)}-\frac{\boldsymbol{\tau}  \times \boldsymbol{b}}{2\pi}\left(\frac{1-2\nu}{4(1-\nu)}\ln{(y_0^2+z_0^2)}+\frac{y_0^2-z_0^2}{4(1-\nu)(y_0^2+z_0^2)}\right)
\end{equation}	
 where the $x_0$ axis is directed along $\boldsymbol{\tau}$, the $z_0$ axis is directed along the vector $\boldsymbol{\tau}  \times \boldsymbol{b}$, and $\nu$ is the Poisson ratio ( $\nu$ =  0.22 for silicon).

 We are only concerned with the $u_x$ component of the $\boldsymbol{u}$ vector. Therefore, for the case considered here the equation \eqref{eq:fov12} simplifies to:
   \begin{equation}
\label{eq:fov13}
	u_x= \frac{a}{\sqrt{2}2\pi}\arctan\frac{z_0}{y_0}
\end{equation}	
for the screw dislocation and to 
     \begin{equation}
\label{eq:fov14}
	u_x= \frac{a}{\sqrt{2}2\pi}\left(\arctan\frac{z_0}{y_0} + \frac{y_0z_0}{4(1-\nu)(y_0^2+z_0^2)}\right)
\end{equation}
for the 60 degree mixed dislocation.
The coordinates ($\textit{$x_0$,$y_0$,$z_0$}$) are related to the coordinates ($\textit{x,y,z}$) via the matrix transformation $\boldsymbol{M}$, where $x_d$, $y_d$ and $z_d$ are positions of origin of the intrinsic coordinate system of dislocation with respect to the origin of the coordinates ($\textit{x,y,z}$).

   \begin{equation}
\label{eq:fov15}
\begin{bmatrix}
x_0\\ 
y_0 \\
z_0
\end{bmatrix}
= \boldsymbol{M}
\begin{bmatrix}
x-x_d\\ 
y-y_d \\
z-z_d
\end{bmatrix}
\end{equation}

where  $\boldsymbol{M}$ = $\begin{bmatrix}
1 & 0 & 0\\ 
0 & -1 & 0 \\
0 & 0 & 1
\end{bmatrix}$ for the screw dislocation case, and $\boldsymbol{M}$ = $\begin{bmatrix}
\frac{1}{2} & -\frac{\sqrt{3}}{2} & 0\\ 
\frac{\sqrt{3}}{2} & -\frac{1}{2} & 0 \\
0 & 0 & 1
\end{bmatrix}$ for the mixed 60 degrees dislocation case.

\begin{figure} [H]
   \begin{center}
   \begin{tabular}{c} 
    \includegraphics[height=4cm]{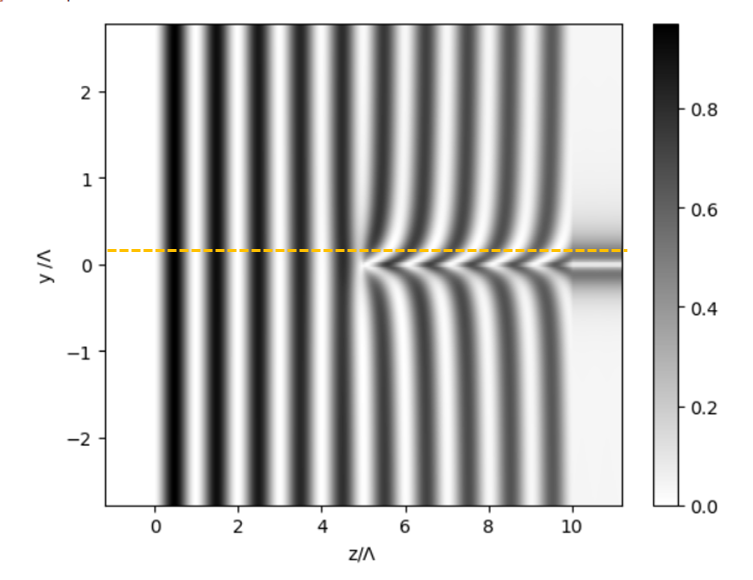}
   \includegraphics[height=4cm, width=5cm]{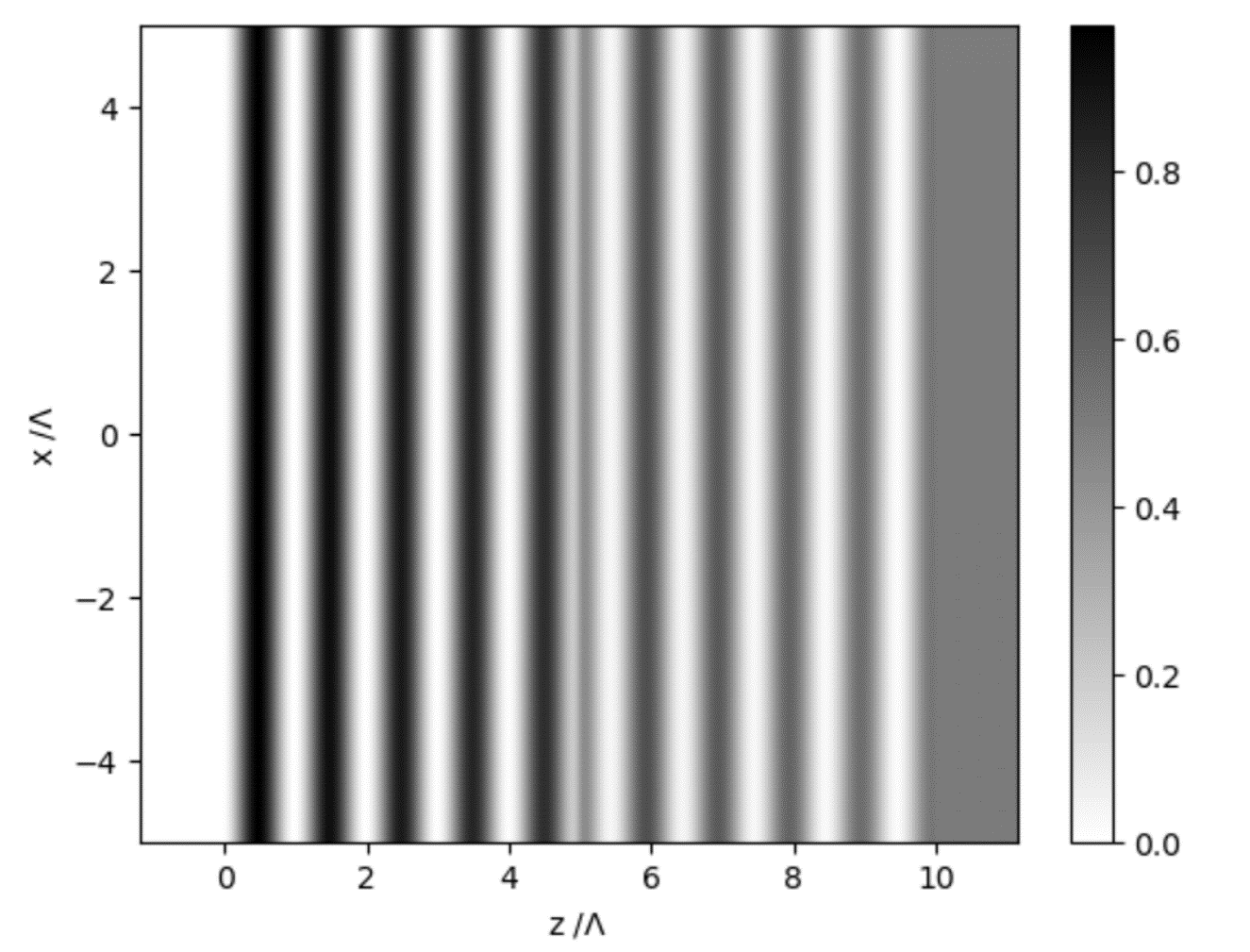}
   \end{tabular}
   \end{center}
   \caption[ScrewDisloc] 
   { \label{fig:ScrewDisloc} 
The intensity of the reflected wave in the Si crystal with a screw dislocation, calculated using the XBPM method, is shown in the yz (left) and xz (right) planes. The cross-section in the xy plane is taken at the position of x, indicated by the dotted line in the left figure.}
   \end{figure}

\begin{figure} [ht]
   \begin{center}
   \begin{tabular}{c} 
    \includegraphics[height=4cm]{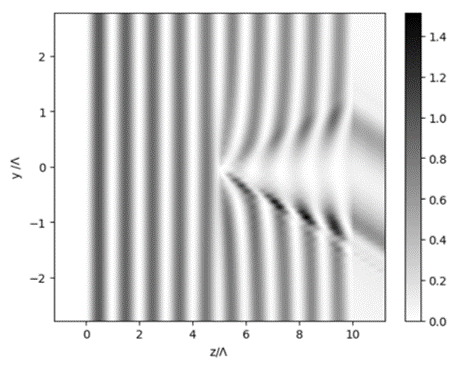}
   \includegraphics[height=4cm, width=5cm]{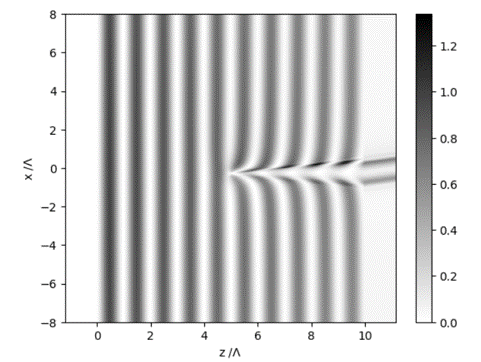}
   \end{tabular}
   \end{center}
   \caption[MixCh] 
   { \label{fig:MixCh} 
Intensity of the reflected wave in the Si crystal with a mixed 60 degrees  dislocation.The results are identical to to the results obtained by the TTE method (Fig. 4(e) in \cite{Besedin2014-bk}). }
   \end{figure}

  Results of XBPM simulations are presented in Fig. \ref{fig:ScrewDisloc} and Fig. \ref{fig:MixCh}. In general, the XBPM results match those obtained by the TTE method by Besedin et al. \cite{Besedin2014-bk}. For example, the intensity distribution inside the crystal, shown on the left side of Fig. \ref{fig:ScrewDisloc}, coincides with the intensity presented in Fig. 4(a) of \cite{Besedin2014-bk}. Similarly, the distribution shown on the left side of Fig. \ref{fig:MixCh} agrees with the intensity pattern presented in Fig. 4(e) of \cite{Besedin2014-bk}. We also noticed, when comparing the results, that the vertical scale of Fig. 4(e) in the work of I. S. Besedin et al. was not consistent with the trigonal coordinate system used, requiring an increase in scale of approximately 25 \%. Only after making this correction did the results obtained by XBPM and TTE agree.


\begin{figure} [H]
   \begin{center}
   \begin{tabular}{c} 
    \includegraphics[height=4cm]{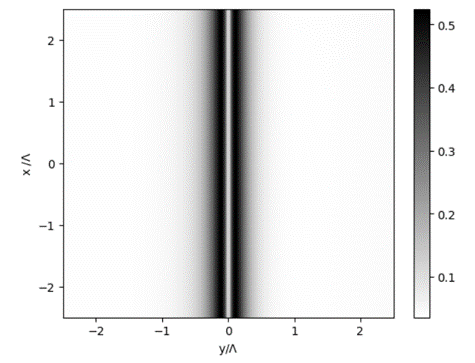}
   \includegraphics[height=4cm, width=5cm]{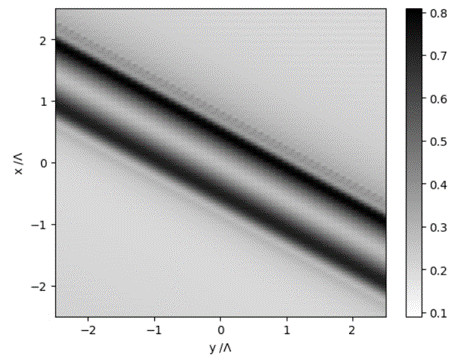}
   \end{tabular}
   \end{center}
   \caption[Topog] 
   { \label{fig:Topog} 
Intensity of the diffracted wave in the xy plane after the beam exits the crystal, left for the screw dislocation and, right for the mixed 60 degrees dislocation.}
   \end{figure}

We also show in Fig. \ref{fig:Topog} the intensity of the diffracted wave in the \textit{xy} plane after the beam leaves the crystal. This corresponds to X-ray topography images taken with a monochromatic beam (Fig. \ref{fig:Topog}). The corresponding Jupyter notebooks and configuration files are:
\begin{itemize}
    \item \texttt{SingleRealization\_Si220\_17p45keVScrewDislChukhovskii.ipynb}
    \item \texttt{Si220\_17p45keVScrewDislChukhovskii.yaml}
\end{itemize}

for the screw dislocation and 
\begin{itemize}
    \item \texttt{SingleRealization\_60degDislocationSi220\_17p45keVChukhovskii.ipynb}
    \item \texttt{Si220\_17p45keV60degDislChukhovskii.yaml}
\end{itemize}
 
for the 60-degree mixed dislocation.

\begin{figure}[H]
   \begin{center}
   \begin{tabular}{c} 
   \includegraphics[width=0.3\linewidth]{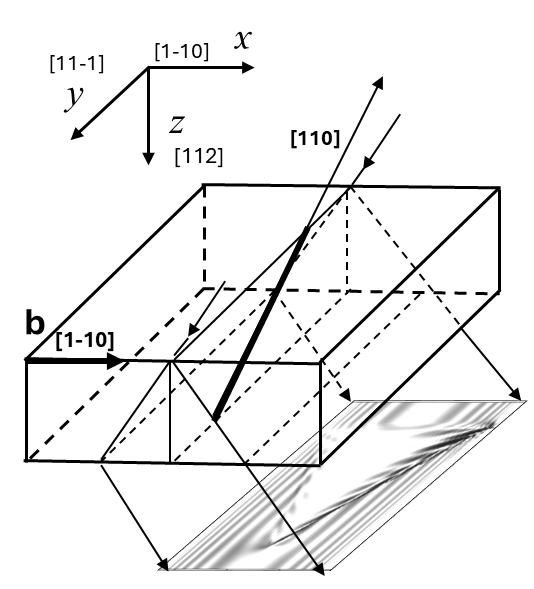}
   \end{tabular}
   \end{center}
   \caption[MixedDisclLang] 
   { \label{fig:MixedDislocLang} 
The geometry of the simulation.}
   \end{figure}

    In the previous examples, the dislocation lines were parallel to the crystal's scattering surface. Now, we will present an example where the dislocation line is not parallel to the surface. We will simulate an X-ray section topography experiment as described in the paper by G. Kowalski et al.\cite{KowalskiGronkowski}. The simulations were performed for the (440) and (220)reflection at 17,450 eV photon energy. A 415 µm thick [112]-oriented silicon single crystal contained a  mixed 60 degrees  [110] dislocation line with a 1/2 [1 -1 0] Burgers vector \textbf{b}. The geometry of the simulation is shown in  Fig. \ref{fig:MixedDislocLang}. The transformation matrix $\boldsymbol{M}$ for this case is given by :  $\boldsymbol{M}$ = $\begin{bmatrix}
0 & -\sqrt{\frac{2}{3}} & -\sqrt{\frac{1}{3}}\\ 
1 & 0 & 0 \\
0 &  \sqrt{\frac{1}{3}} & -\sqrt{\frac{2}{3}}
\end{bmatrix}$. In the X-ray section topography method, the incoming X-ray beam is shaped by a slit with a width on the order of micrometers. I our case we have applied 5 $\mu$m wide slit.

        \begin{figure}[H] 
   \begin{center}
   \begin{tabular}{c} 
   \includegraphics[width=0.3\linewidth]{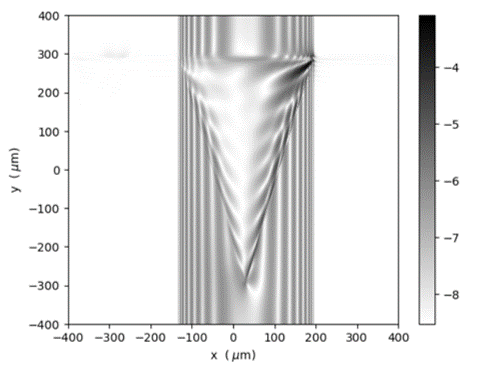}
   \includegraphics[width=0.3\linewidth,height=4cm, width=5cm]{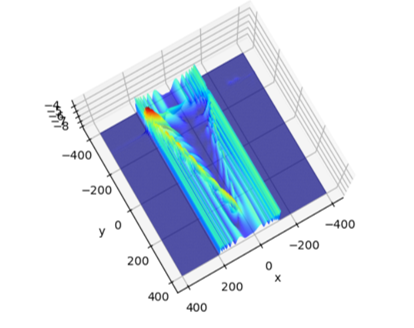}
   \end{tabular}
   \end{center}
   \caption[MixedDisclLangSi440] 
   { \label{fig:MixedDisclLangSi440} 
Simulation of a topography of a mixed 60 deg dislocation in a Si crystal by the Lang method for (440) reflection. The intensity scale is logarithmic}
   \end{figure}

\begin{figure}[H]
   \begin{center}
   \begin{tabular}{c} 
   \includegraphics[width=0.3\linewidth]{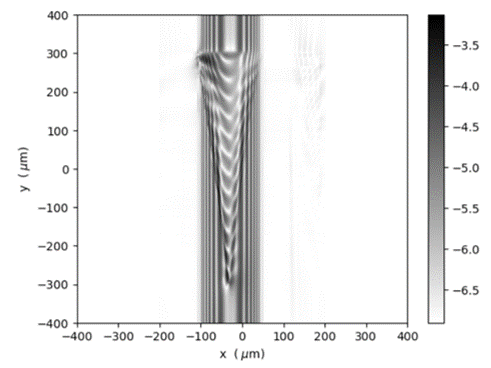}
   \includegraphics[width=0.3\linewidth,height=4cm, width=5cm]{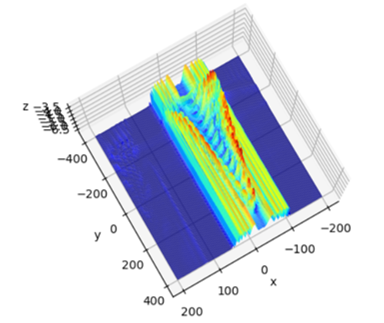}
   \end{tabular}
   \end{center}
   \caption[MixedDisclLangSi220] 
   { \label{fig:MixedDisclLangSi220} 
Simulation of a topography of a 60 deg dislocation in a Si crystal by the Lang method for (220) reflection. The intensity scale is logarithmic}
   \end{figure}

     The results of the simulations, presented in Fig. \ref{fig:MixedDisclLangSi440} and Fig. \ref{fig:MixedDisclLangSi220}, are in qualitative agreement with those presented in the work of G. Kowalski et al \cite{KowalskiGronkowski}. Unfortunately, we were unable to make a quantitative comparison, as the authors of the cited work did not provide sufficient information regarding intensity scaling in the presented figures.
     
     The corresponding Jupyter notebooks and configuration files are:
\begin{itemize}
    \item \texttt{SingleRealization\_Si440\_60deg\_DislocationGronkowski.ipynb}
    \item \texttt{Si440\_17p45keVDislk60degGronkowski.yaml}
\end{itemize}
 
 for the (400)  reflection,

and 

\begin{itemize}
 \item \texttt{SingleRealization\_Si220\_60deg\_DislocationGronkowski.ipynb} and

\item \texttt{Si220\_17p45keVDislk60degGronkowski.yaml} 
\end{itemize}

for the (200)  reflection.    
\section{crystals with arbitrary shapes and strain distributions}

Our FFT BPM method simplifies simulations compared to the TTE method by eliminating the need to solve boundary condition problems. Therefore, it is well-suited for simulating scattering from finite crystals with arbitrary shapes and strain distributions. All that is required are maps of the crystal boundaries and the deformation vector component $u_x$, which is perpendicular to the crystal planes.

We will illustrate this approach with several examples presented below.

First, we apply our method to simulate reflection from a Si(004) rectangular crystal with dimensions 100~$\mu$m~$\times$~300~$\mu$m at 12~keV photon energy, an example also simulated in the article \textit{``X-ray dynamical diffraction from single crystals with arbitrary shape and strain field: A universal approach to modeling''} \cite{PhysRevB.89.014104}. The crystal boundary map is generated using the \texttt{geometry-finite Si crystal-2D.ipynb} Jupyter notebook.

    \begin{figure} [ht]
   \begin{center}
   \begin{tabular}{c} 
    \includegraphics[height=3.5cm, width=5cm]{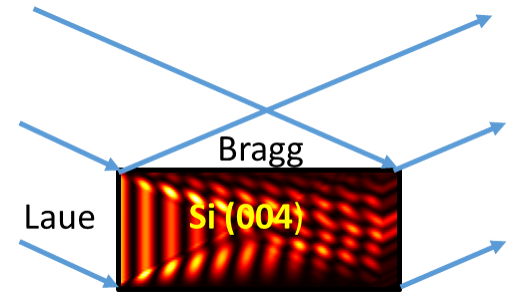}
   \includegraphics[height=3cm]{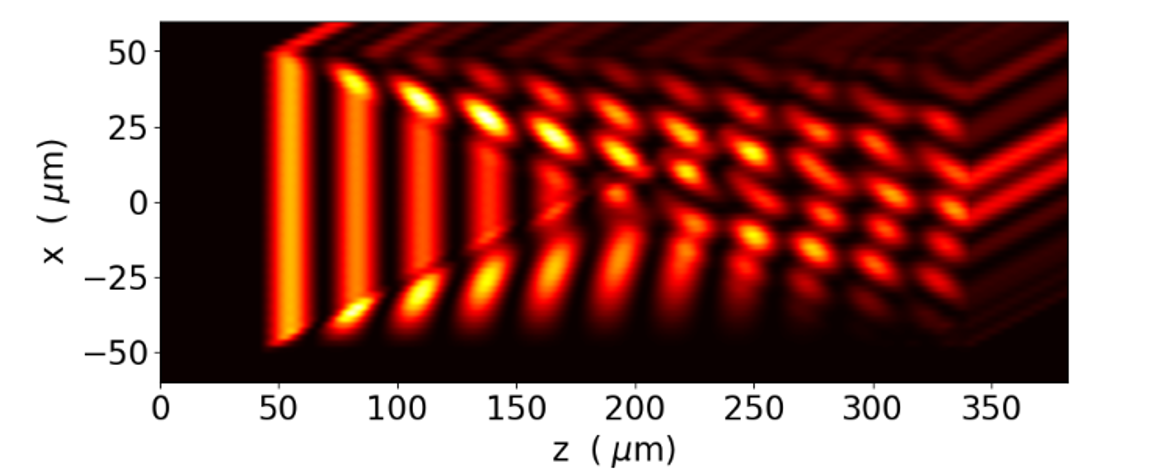}
   
   \end{tabular}
   \end{center}
   \caption[finitCryst] 
   { \label{fig:finitCryst} 
Reflection from a Si(004) rectangular crystal with dimensions  100 $\mu$m x 300 $\mu$m at 12 keV photon energy.  Pendellösung fringes are visible at the crystal's entrance. A more complicated diffraction patter develops for the mixed Laue - Bragg case. }
   \end{figure}

The result of the simulation, which is identical to the one  presented in \cite{PhysRevB.89.014104},  is shown in Fig. \ref{fig:finitCryst}.  Pendellösung fringes are visible at the crystal's entrance. A more complicated diffraction patter develops for the mixed Laue - Bragg case.
The corresponding Jupyter notebooks and configuration files are:
\begin{itemize}
  \item \texttt{SingleRealization\_Si400\_12keV\_FiniteCrystal-2D.ipynb}

 \item \texttt{Si400\_12keV\_finite\_Crystal\_2D.yaml}
\end{itemize}

    \begin{figure} [H]
   \begin{center}
   \begin{tabular}{c} 
    \includegraphics[height=8cm, width=12cm]{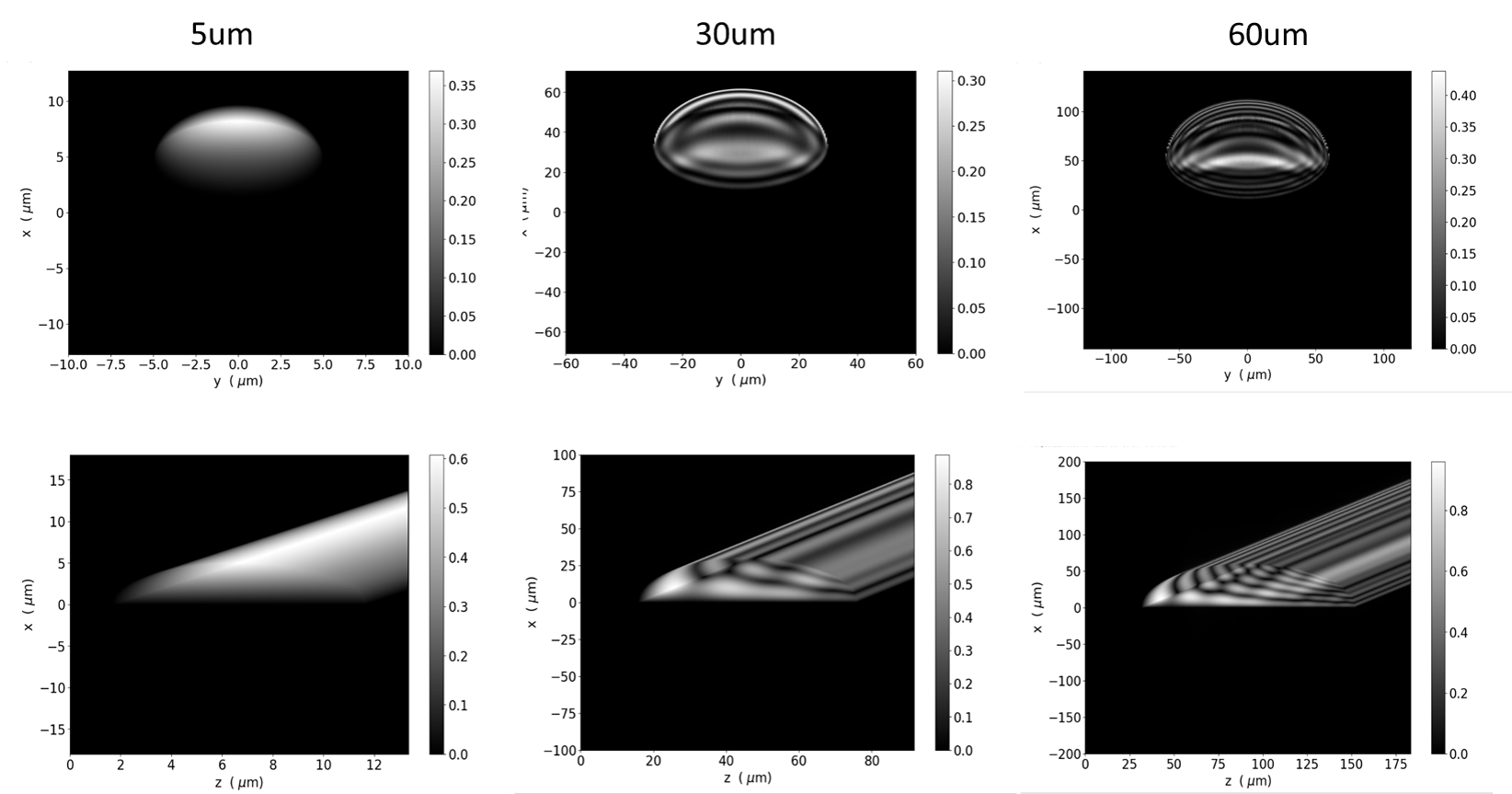}
  
   \end{tabular}
   \end{center}
   \caption[Semispheres] 
   { \label{fig:Semispheres} 
Reflection from a C(400) semisphere crystals with dimensions 5 $\mu$m, 30 $\mu$m and 60 $\mu$m at 9.8 keV photon energy. Cross-section of the reflected wave in the $xz$-plane at y=0 (lower row) and reflected  wave intensity in the $xy$-plane after exiting the crystal. A complicated diffraction patter develops when the crystal size is larger than extinction length. } 
   \end{figure}
Next, we present simulation results for semi-spherical C(400) crystals with radii of 5 $\mu$m, 30 $\mu$m, and 60 $\mu$m at 9.8 keV photon energy, shown in Fig. \ref{fig:Semispheres}. These results highlight the necessity of accounting for dynamical diffraction effects as the crystal size approaches or exceeds the extinction length. For a radius of 5 $\mu$m, the simulated intensity closely resembles predictions from kinematic theory. However, for larger radii, the reflected wave intensity exhibits a complex pattern due to dynamical effects.

The Jupyter notebooks for running the simulation, generating the crystal boundary map, and the configuration file for a radius of 10~$\mu$m are:

\begin{itemize}
    \item \texttt{SingleRealization\_C400-hemi-sphere.ipynb}
    \item \texttt{geometry-hemi-sphere.ipynb}
    \item \texttt{C400\_9p8keV\_Laue-hemi-sphere.yaml}
\end{itemize}

\begin{figure} [H]
   \begin{center}
   \begin{tabular}{c} 
    \includegraphics[height=12cm, width=14cm]{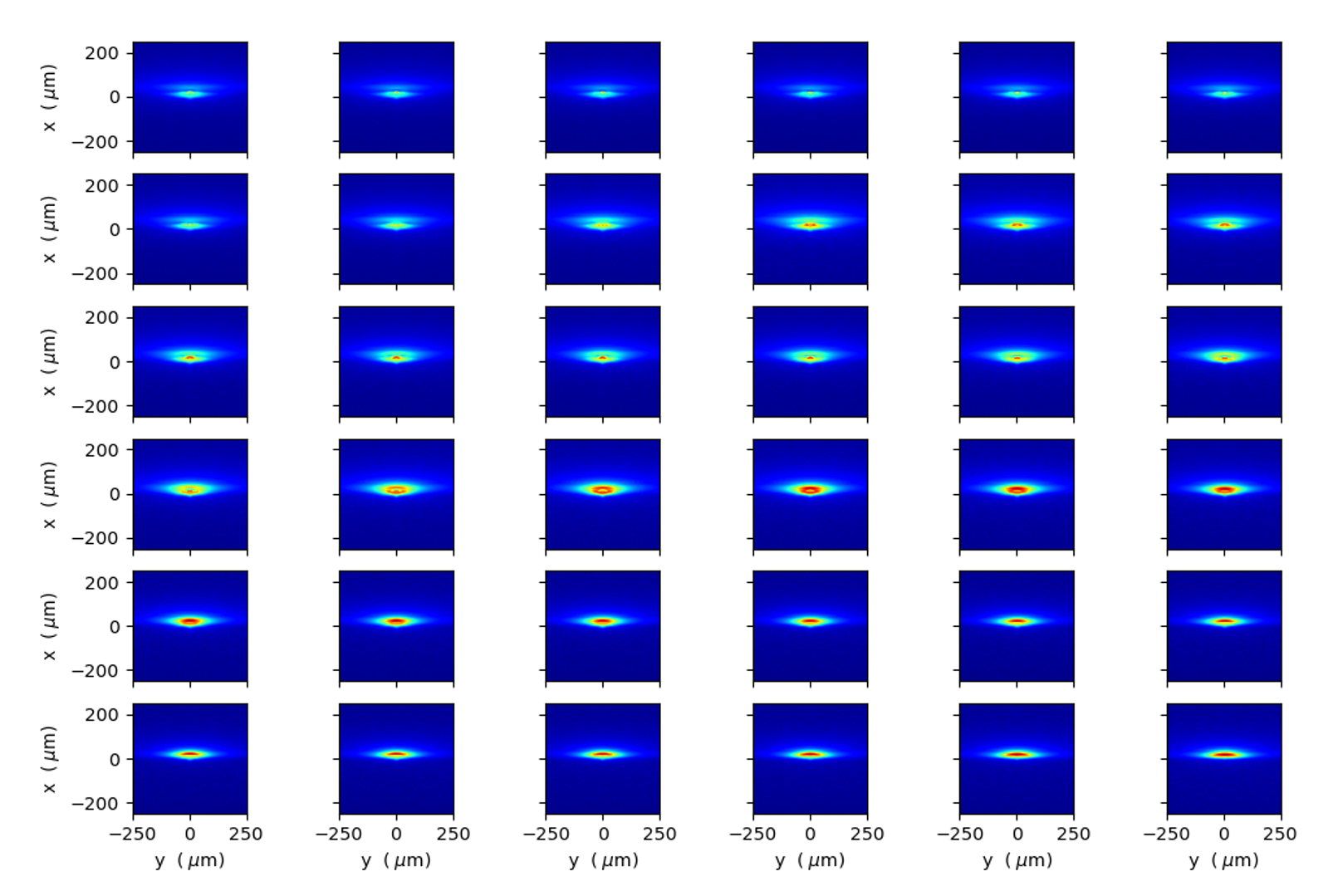}
   \end{tabular}
   \end{center}
   \caption[hemi-far-AngScan] 
   { \label{fig:hemi-far-AngScan} 
   Far-field images simulated 2~m from the crystal, obtained from an angular scan of $\pm 10~\mu$rad.} 
\end{figure}

\begin{figure} [H]
   \begin{center}
   \begin{tabular}{c} 
    \includegraphics[height=12cm, width=14cm]{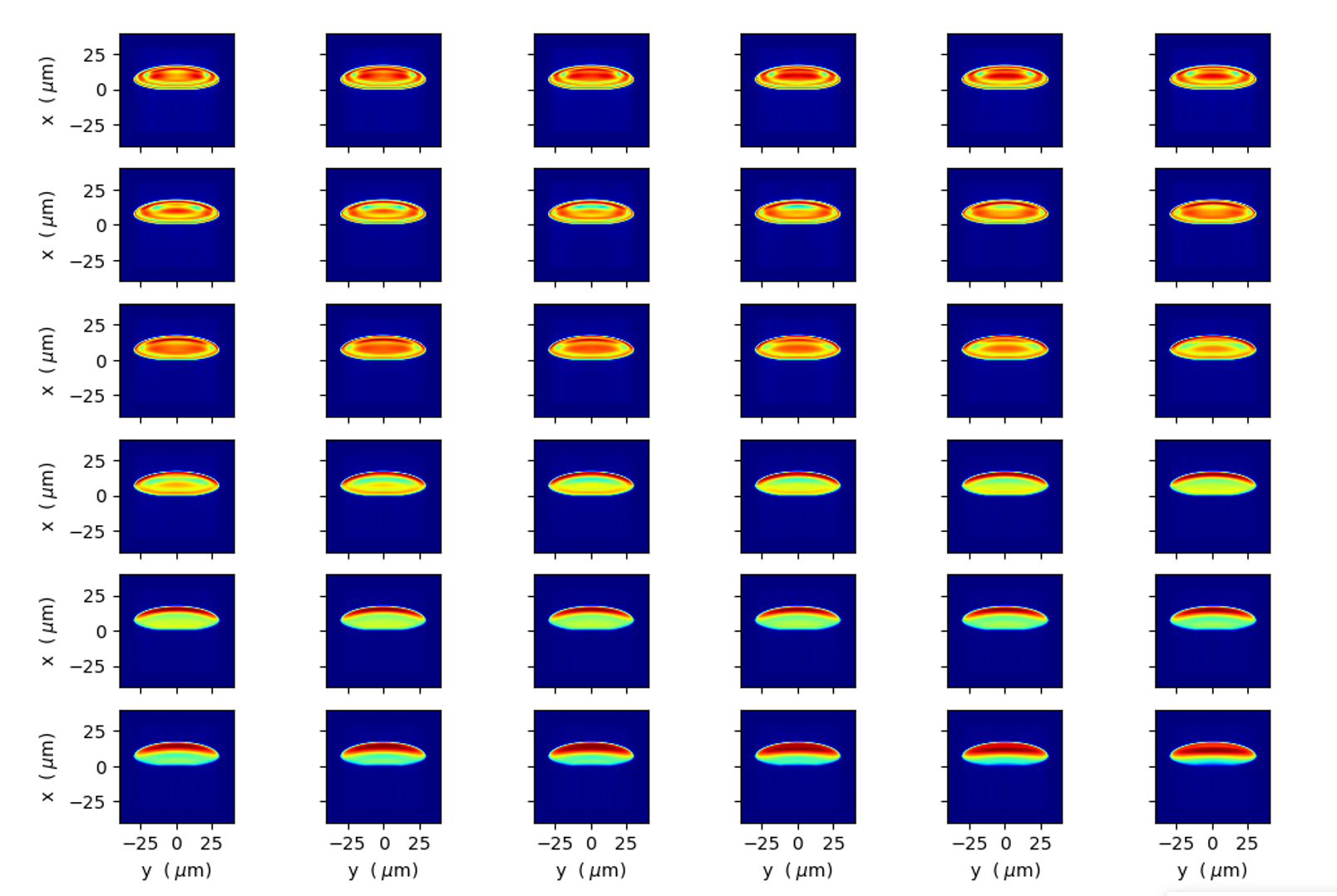}
   \end{tabular}
   \end{center}
   \caption[hemi-near-AngScan] 
   { \label{fig:hemi-near-AngScan} 
   Near-field images obtained from an angular scan of $\pm 10~\mu$rad.} 
\end{figure}

Additionally, we provide a Python script for simulating an angular scan of the hemisphere with a 10~$\mu$m radius as it is rocked in the $xz$-plane. Far-field images simulated 2~m from the crystal over an angular range of $\pm 10~\mu$rad are shown in Fig.~\ref{fig:hemi-far-AngScan}. For reference, Fig.~\ref{fig:hemi-near-AngScan} displays the corresponding near-field images.

The corresponding Python code for parallel computing, the configuration file, and the Jupyter notebook for displaying the results are as follows:

\begin{itemize}
    \item \texttt{run\_parallel\_angleC400\_9p8keV\_Laue\_Asym-sphere.py} 
    \item \texttt{C400\_9p8keV\_Laue\_Asym-sphere-par.yaml}
    \item \texttt{process-parallel-data-angleC400\_9p8keV\_Laue\_Asym-sphere.ipynb} 
\end{itemize}

\begin{figure} [H]
   \begin{center}
   \begin{tabular}{c} 
    \includegraphics[height=3.5cm, width=5cm]{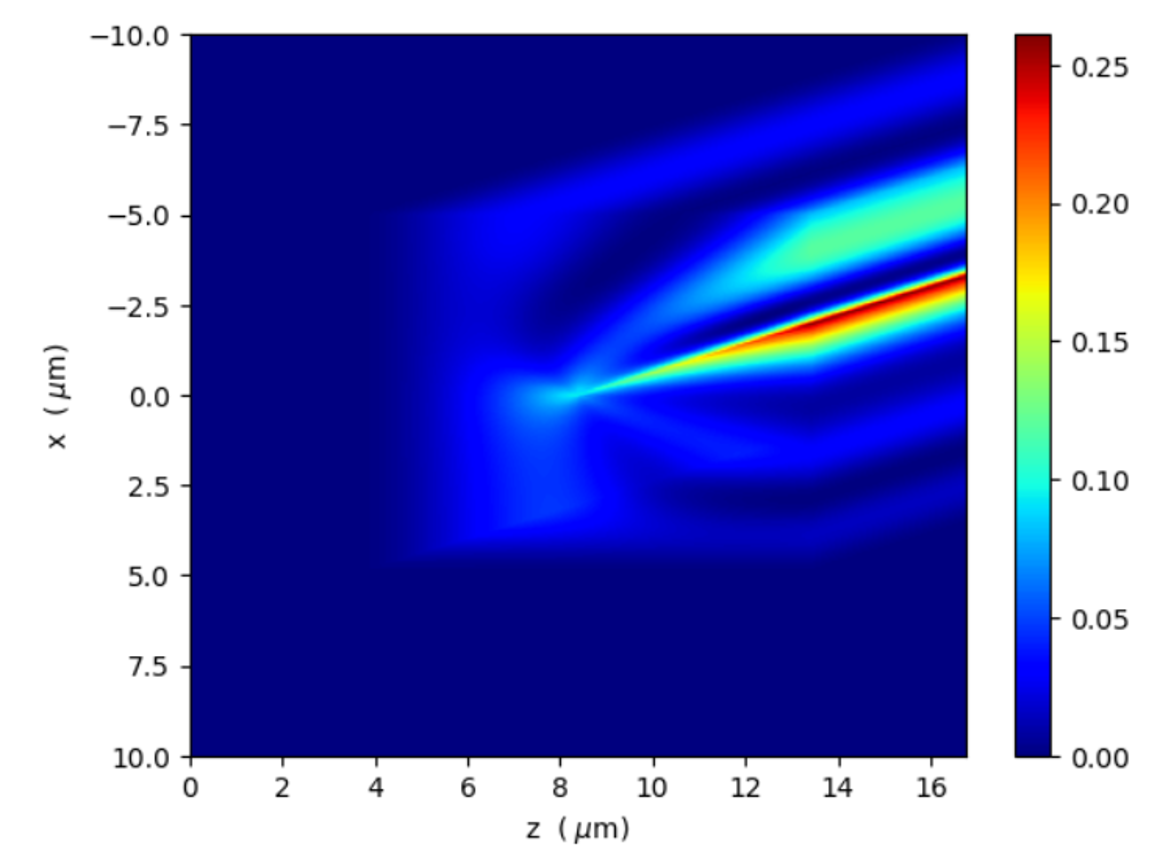}
    \includegraphics[height=3.5cm, width=5cm]{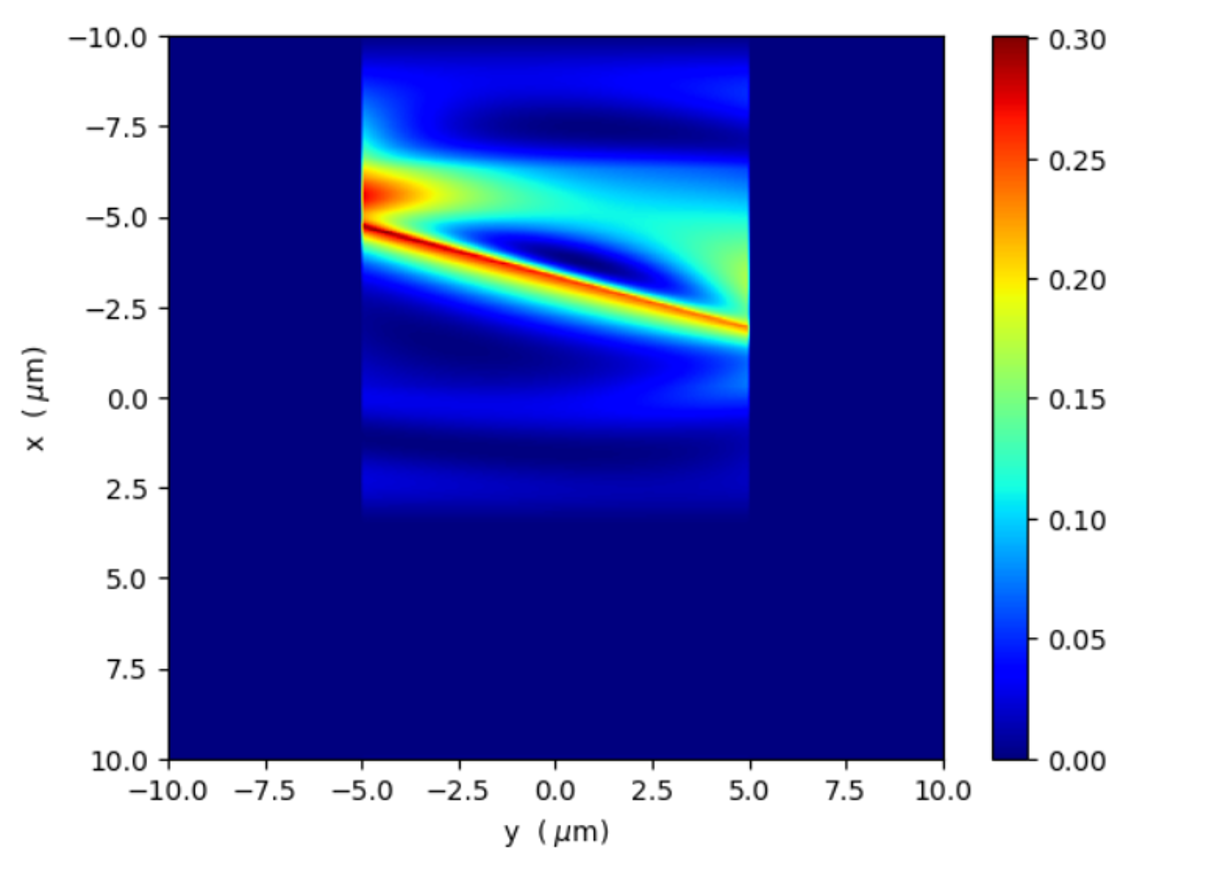}
    \includegraphics[height=4cm, width=6cm]{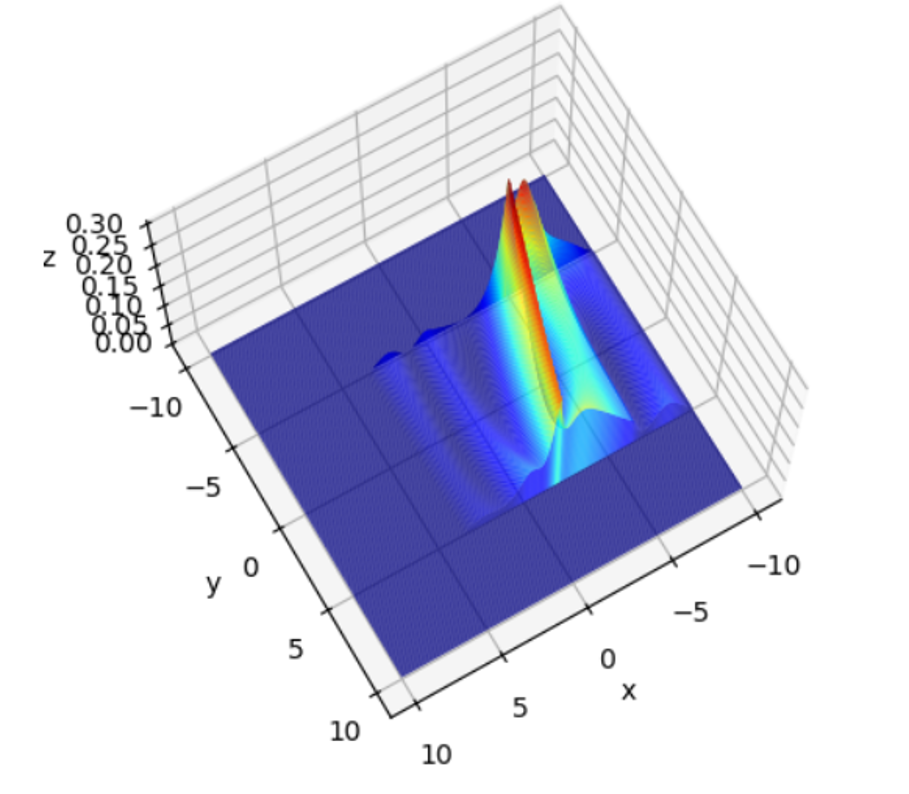}
    \\
    \includegraphics[height=3.5cm, width=5cm]{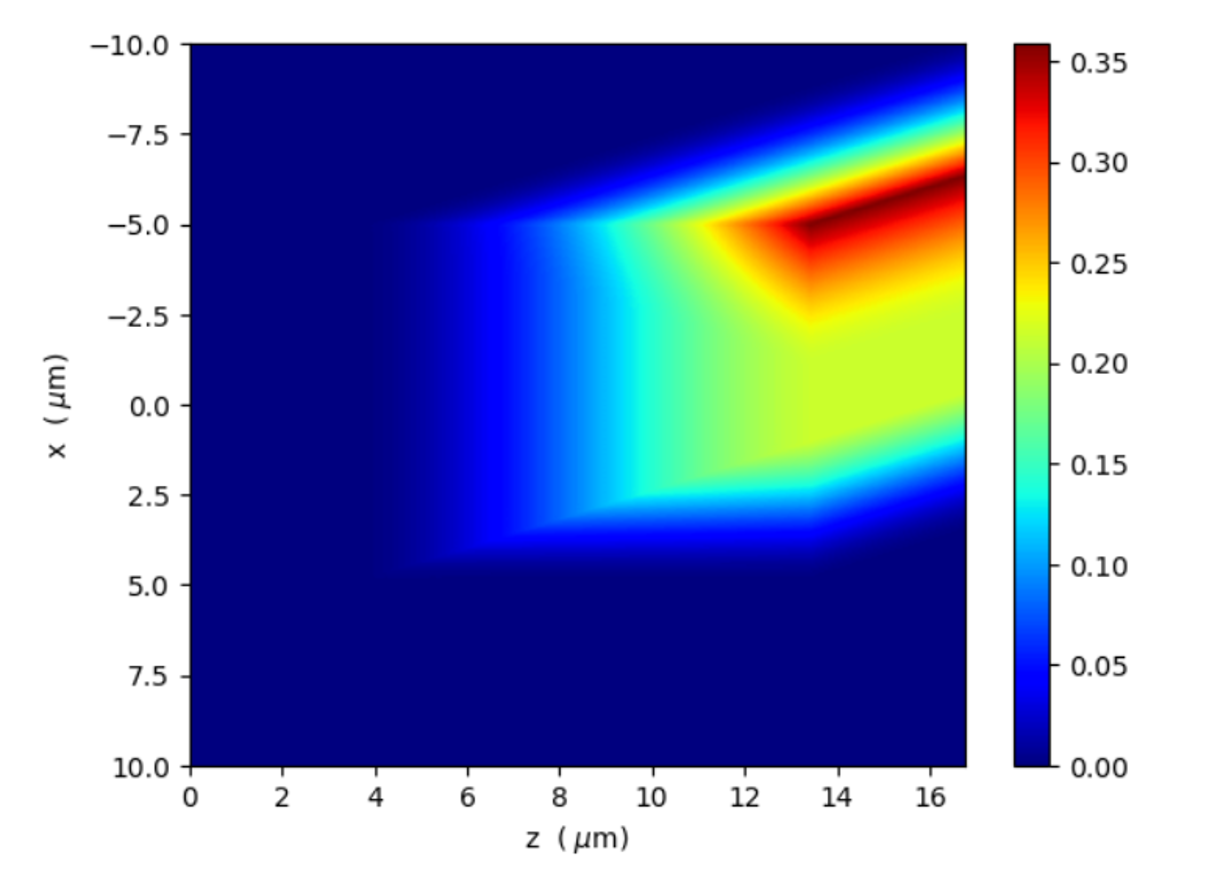}
    \includegraphics[height=3.5cm, width=5cm]{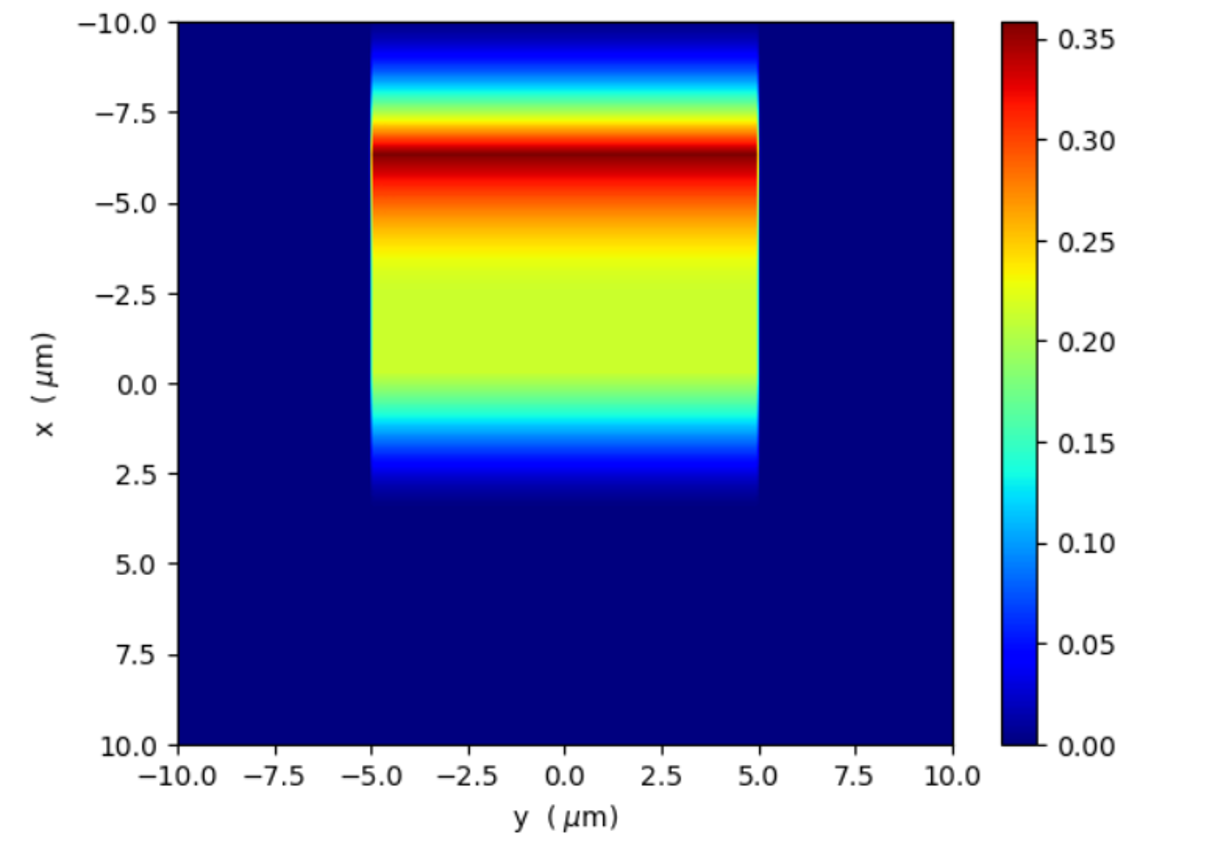}
    \includegraphics[height=4cm, width=6cm]{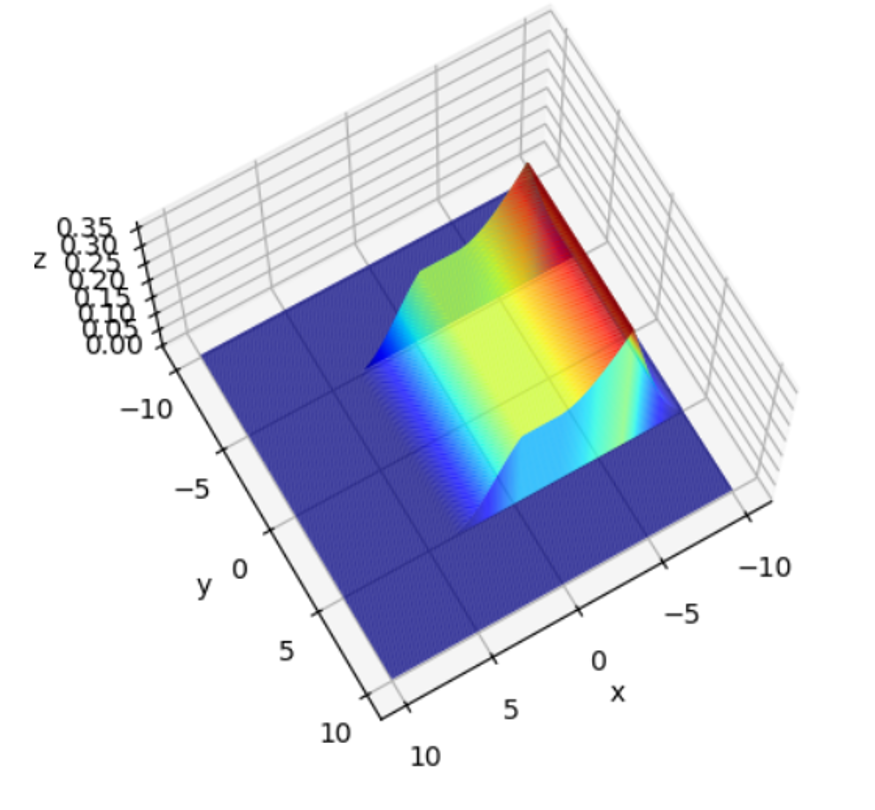}

   \end{tabular}
   \end{center}
   \caption[FiniteCrystalDisloc] 
   { \label{fig:FiniteCrystalDisloc} 
Si(440) reflection from a 10~$\mu$m silicon cube crystal at a photon energy of 17,450~eV. The crystal is deformed by a mixed 60\textdegree{} [110] dislocation with a $\frac{1}{2}[1\,\overline{1}\,0]$ Burgers vector. The top left panel shows the cross-section of the reflected wave in the $xz$-plane at $y=0$, while the top center and right panels display the reflected wave intensity in the $xy$-plane after exiting the crystal. For comparison, the lower row presents simulation results for the same crystal without a dislocation. } 
   \end{figure}

\begin{figure} [H]
   \begin{center}
   \begin{tabular}{c} 
    \includegraphics[height=3.5cm, width=5cm]{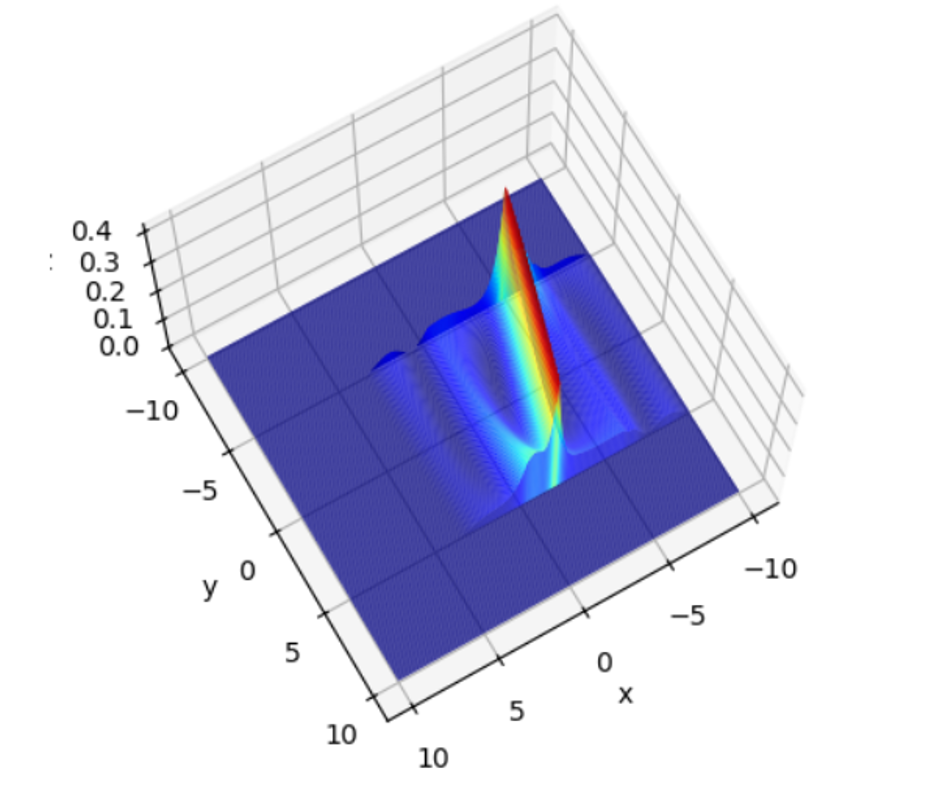}
    \includegraphics[height=3.5cm, width=5cm]{FiniteCrystalDislocXY3D.png}
    \includegraphics[height=3.5cm, width=5cm]{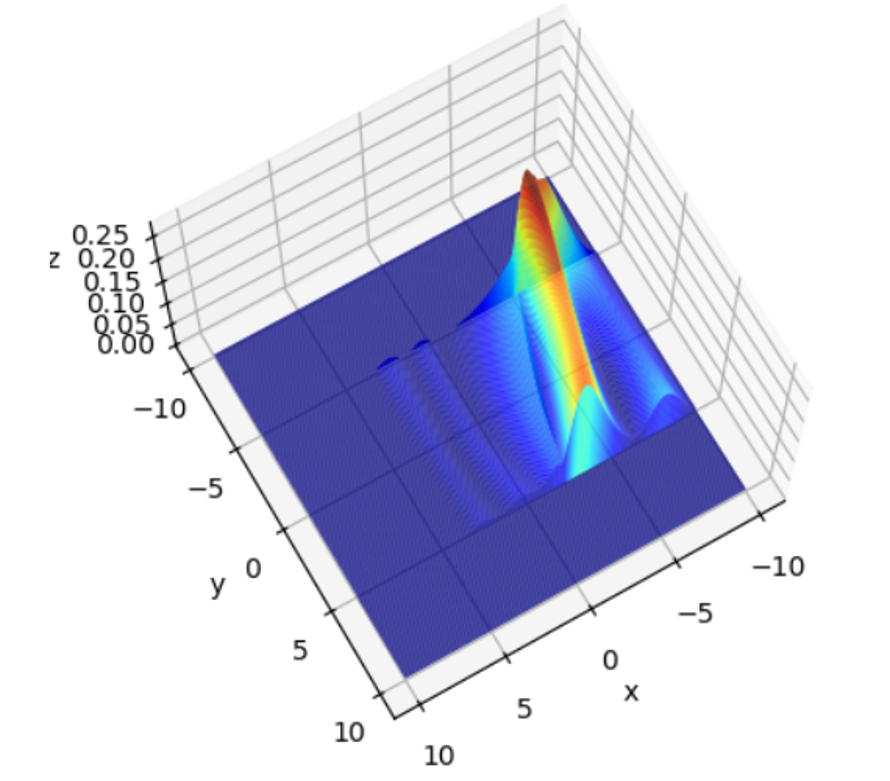}
  
   \end{tabular}
   \end{center}
   \caption[FiniteCrystalDislocDiffAng] 
   { \label{fig:FiniteCrystalDislocDiffAng} 
Reflected wave intensity in the $xy$-plane after exiting the crystal for different incidence angles: at the Bragg angle (center), $-3~\mu\text{rad}$ below the Bragg angle (left), and $+3~\mu\text{rad}$ above the Bragg angle (right).} 
 \end{figure}

In the final example in this section, we simulate Si(440) reflection from a 10~$\mu$m silicon cube crystal deformed by a mixed 60° [110] dislocation with a $\frac{1}{2}$[1 -1 0] Burgers vector at a photon energy of 17,450~eV. The simulation results are shown in Fig.~\ref{fig:FiniteCrystalDisloc} and Fig.~\ref{fig:FiniteCrystalDislocDiffAng}. We highlight the numerical efficiency of our approach: the simulations, performed on a $300 \times 200 \times 300$ rectangular grid, ran on a single node of the Perlmutter high-performance machine \cite{Perlmutter} and took approximately 13 seconds.

The Jupyter notebooks for running the simulation, generating the crystal boundary map, and the configuration file for a 10~$\mu$m silicon cube crystal deformed by a mixed 60° [110] dislocation are:

\begin{itemize}
    \item \texttt{SingleRealizationFiniteCrystal\_Si440\_60deg\_Dislocation.ipynb}
    \item \texttt{geometry-finite Si440crystal-60degDislocation.ipynb}
    \item \texttt{Si440\_17p45keVDislk60degGronkowskiFiniteCrystal3D.yaml}
\end{itemize}

\begin{figure} [H]
   \begin{center}
   \begin{tabular}{c} 
    \includegraphics[height=12cm, width=14cm]{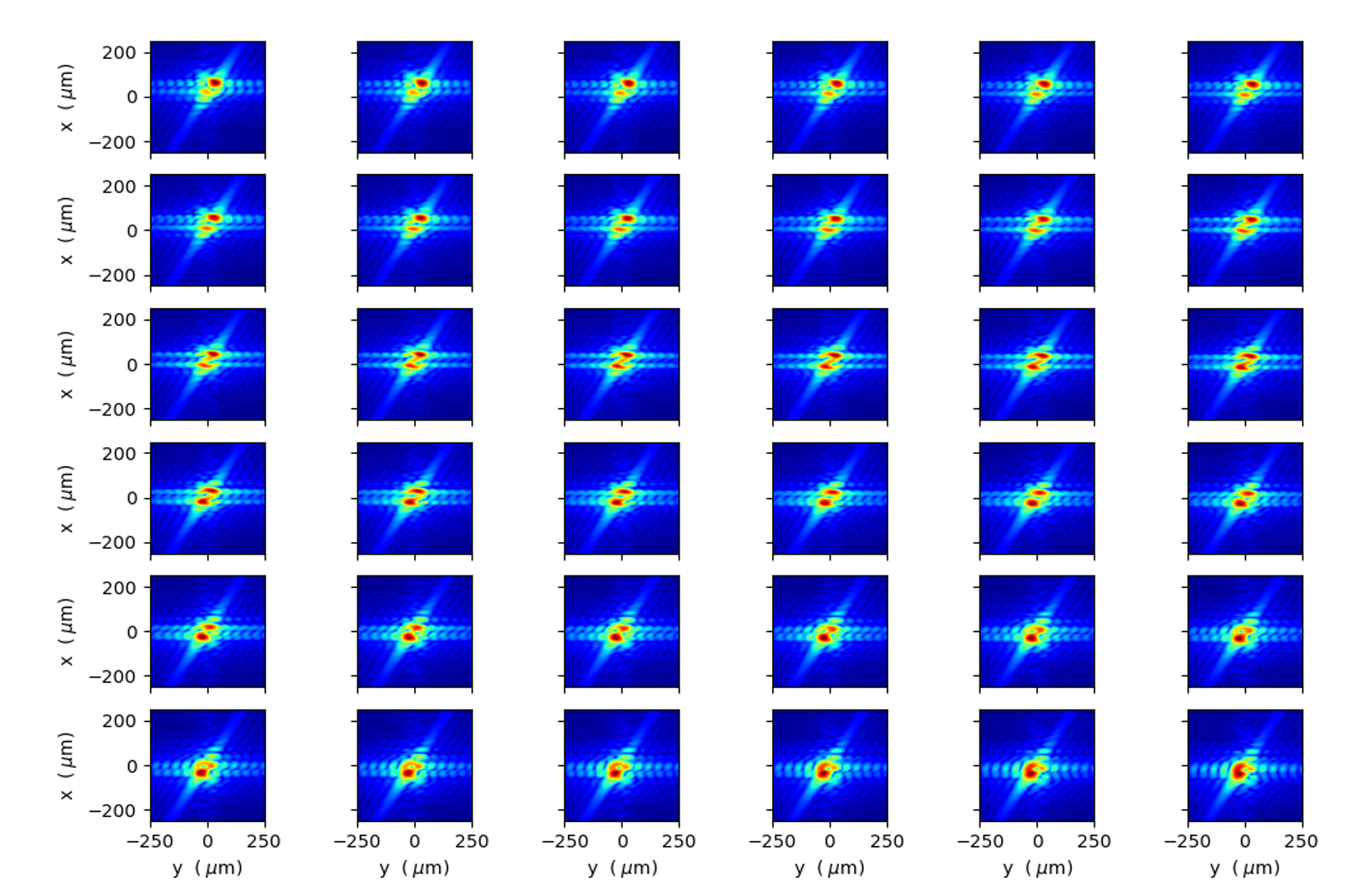}
   \end{tabular}
   \end{center}
   \caption[AngScan] 
   { \label{fig:AngScan} 
   Far-field images simulated 4~m from the crystal, obtained from an angular scan of $\pm 10~\mu$rad.} 
\end{figure}

\begin{figure} [H]
   \begin{center}
   \begin{tabular}{c} 
    \includegraphics[height=12cm, width=14cm]{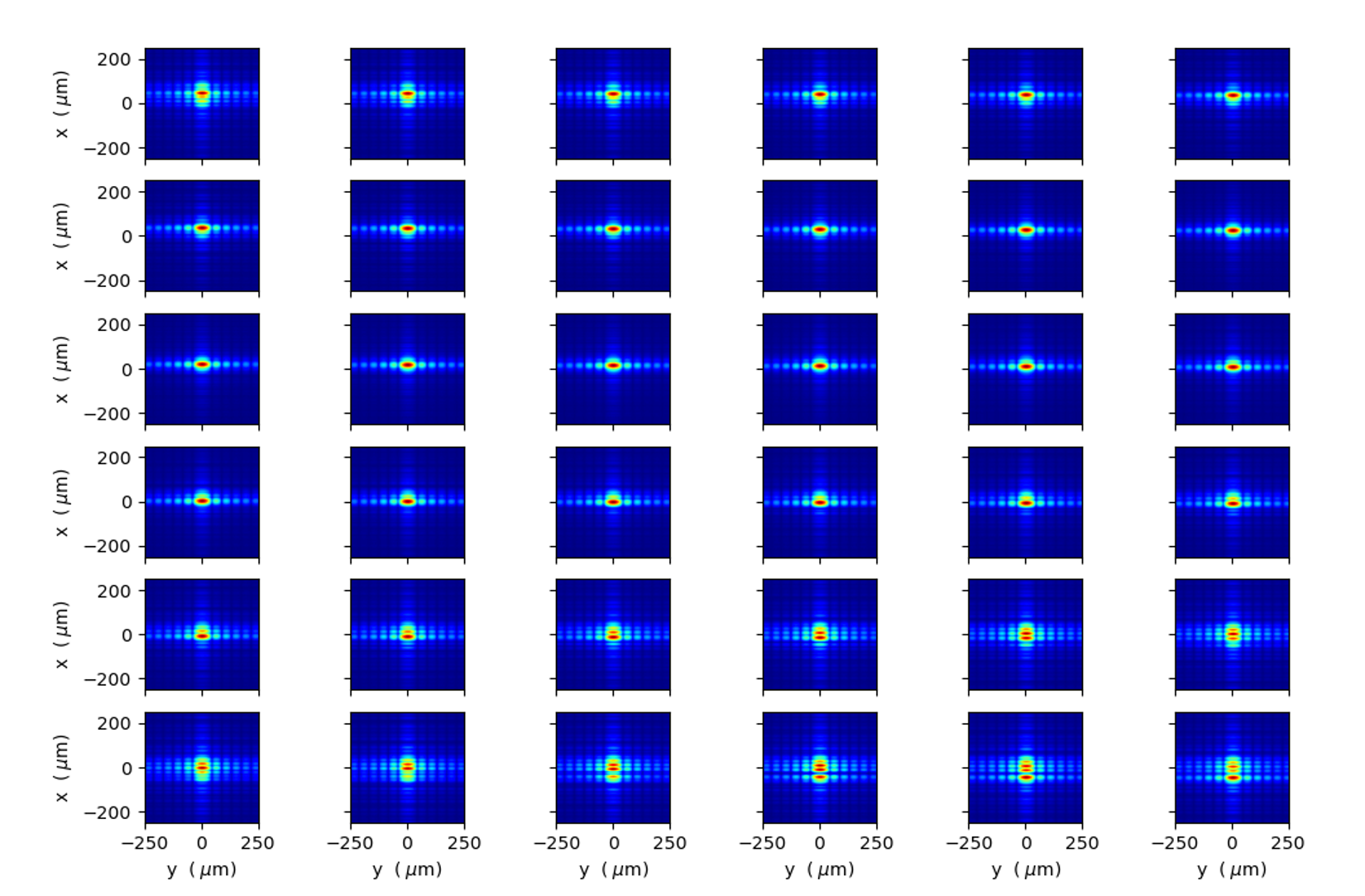}
   \end{tabular}
   \end{center}
   \caption[AngScanNoDisl] 
   { \label{fig:AngScanNoDisl} 
   Far-field images simulated 4~m from the crystal without dislocation, obtained from an angular scan of $\pm 10~\mu$rad.} 
\end{figure}
We also provide a Python script that simulates an angular scan as the crystal is rocked in the $xz$-plane. Far-field images simulated 4~m from the crystal over an angular range of $\pm 10~\mu$rad are shown in Fig.~\ref{fig:AngScan}. For comparison, Fig.~\ref{fig:AngScanNoDisl} presents similar images simulated for the same crystal without the dislocation. 

The corresponding Python code for parallel computing, the configuration file and the Jupyter notebook for displaying the results are as follows:

\begin{itemize}
    \item \texttt{run\_parallel\_angle\_scanFiniteCrystal\_Si440\_60deg\_Dislocation.py} 
    \item \texttt{Si440\_17p45keVDislk60degGronkowskiFiniteCrystal3Dpar.yaml}
    \item \texttt{process-parallel-data-angleFiniteCrystal\_Si440\_60deg\_Dislocation.ipynb} 
    
\end{itemize}

We note that FFT BPM simulations can be particularly useful in the analysis of Bragg coherent X-ray diffractive imaging (CXDI) data, especially when the crystal size approaches the extinction length. Because of the simplicity and computational efficiency of the FFT BPM method, it could be integrated into iterative algorithms to retrieve the shape and strain distribution of the investigated crystal.

 \section{Time-Dependent Simulation of Reflection from Distorted Crystals}

The following section provides an overview of the Jupyter notebooks, configuration files, and Python scripts available in the public repository that were used for the time-dependent simulations described in Ref.~\cite{Krzywinski2022}. The theoretical background and detailed simulation conditions are presented in that reference. Each example in the repository corresponds directly to the figures in Ref.~\cite{Krzywinski2022}, allowing users to reproduce and extend the published results.

\subsection{Benchmarking of the TD FFT BPM Method}

\begin{figure} [H]
   \begin{center}
   \begin{tabular}{c} 
    \includegraphics[height=6cm, width=8cm]{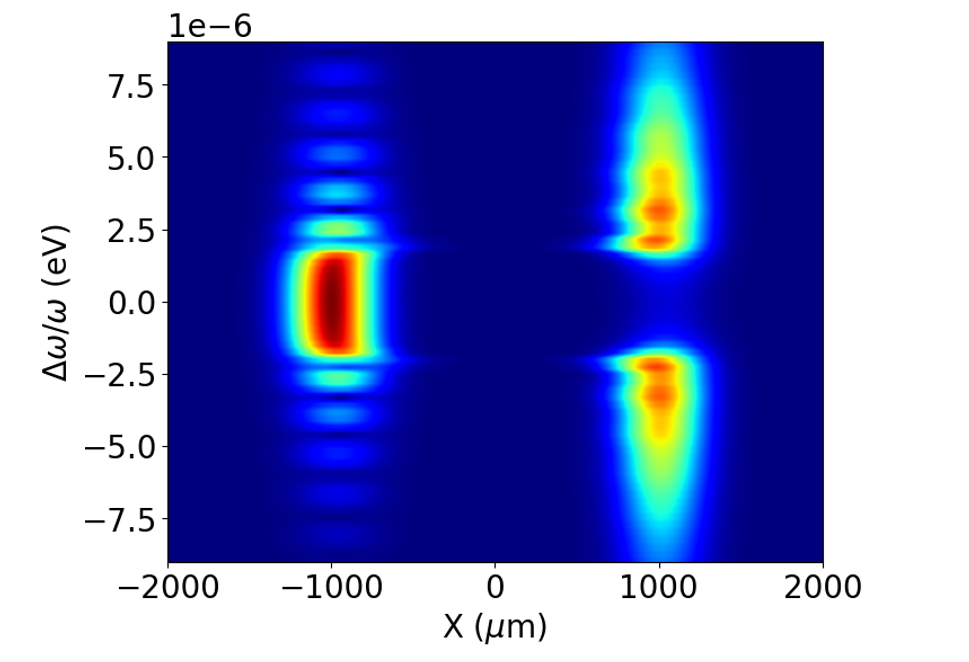}
    \includegraphics[height=6cm, width=8cm]{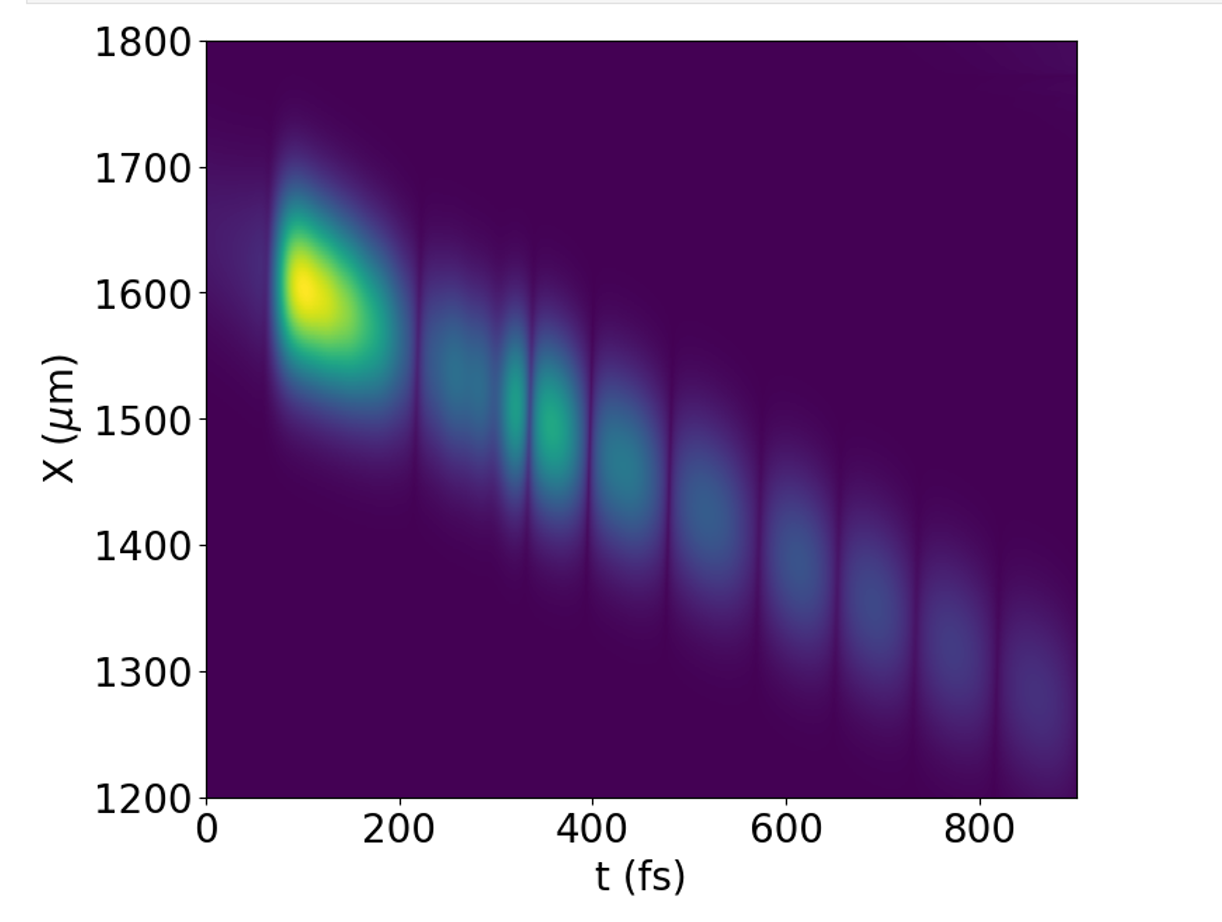}
   \end{tabular}
   \end{center}
   \caption[OmegaX] 
   { \label{fig:OmegaX} 
Spatio-spectral intensity profiles of the reflected and transmitted beams for a 50-µm-thick C*(333) diamond crystal at \(y=0\) using an incident Gaussian beam with \(t = 10\,\mathrm{fs}\), \(W_0 = 250\,\mu\mathrm{m}\), and \(\omega_0 = 12.8\,\mathrm{keV}\) (left), and spatio-temporal intensity profiles of the Bragg-diffracted beam computed with \(W_0 = 50\,\mu\mathrm{m}\) while keeping all other parameters the same (right); color-scale units are arbitrary.} 
\end{figure}

The first set of examples benchmarks the time-dependent Fast Fourier Transform Beam Propagation Method (TD~FFT~BPM) against the analytical solution for perfect crystals presented by Shvyd’ko and Lindberg~\cite{PhysRevSTABShvyd'koLindberg}. The left and right panels in Fig.~\ref{fig:OmegaX} correspond to Figs.~5 and~6 in Ref.~\cite{Krzywinski2022}, respectively; note that in the caption of Fig.~5 in Ref.~\cite{Krzywinski2022}, the beam waist \(W_0\) was mistakenly listed as \(50\,\mu\mathrm{m}\), whereas the correct value is \(250\,\mu\mathrm{m}\).

The simulations shown in Fig.~\ref{fig:OmegaX} were produced using the following Python scripts, configuration files, and Jupyter notebooks:

\textbf{Left panel:}
\begin{itemize}
    \item \texttt{run\_parallel-omegaC333-250um.py}
    \item \texttt{C333\_Omega12p8keV\_waist250um.yaml}
    \item \texttt{process-parallel-data-omegaC333-250.ipynb}
\end{itemize}

\textbf{Right panel:}
\begin{itemize}
    \item \texttt{run\_parallel-omegaC333-50um.py}
    \item \texttt{C333\_Omega12p8keV\_waist50um.yaml}
    \item \texttt{process-parallel-data-omegaC333-50.ipynb}
\end{itemize}

As noted in Ref.~\cite{Krzywinski2022}, the FFT-BPM method shows perfect agreement with the theoretical calculations of Shvyd’ko and Lindberg (2012).

\subsection{Reflection from a Deformed Crystal}

The second set of examples demonstrates time-dependent reflection from a deformed crystal. These simulations correspond to Figs.~7,~8 and 10 in Ref.~\cite{Krzywinski2022}. The examples show how to include deformation fields within the FFT BPM framework to model the resulting time-dependent diffraction response.
\begin{figure} [H]
   \begin{center}
   \begin{tabular}{c} 
       \includegraphics[height=4cm, width=5.5cm]{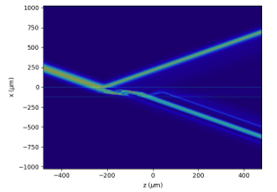}
    \includegraphics[height=4cm, width=5.5cm]{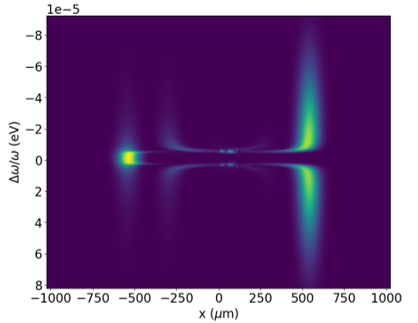}
    \includegraphics[height=4cm, width=5.5cm]{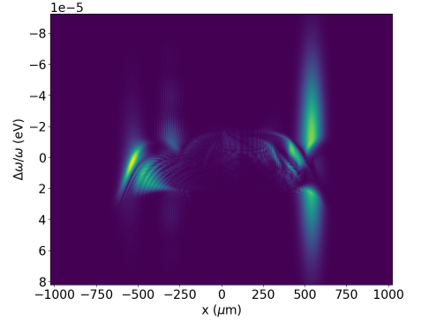}
   \end{tabular}
   \end{center}
   \caption[TDsf-Deformed] 
      { \label{fig:TDsf-Deformed}
      2D visualization of Bragg reflection from the deformed crystal at the central XFEL frequency (left), and spatio-spectral intensity profiles of the Bragg-diffracted beam at \(y = 0\) simulated with the FFT-BPM method for the ideal crystal (center) and for the deformed crystal (right); color-scale units are arbitrary.}
 
\end{figure}

\begin{figure} [H]
   \begin{center}
   \begin{tabular}{c} 

    \includegraphics[height=4cm, width=5.5cm]{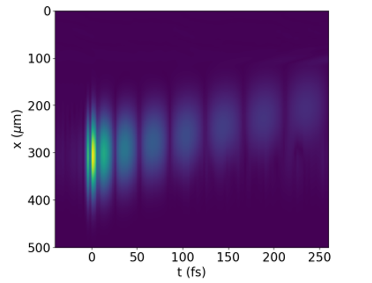}
    \includegraphics[height=4cm, width=5.5cm]{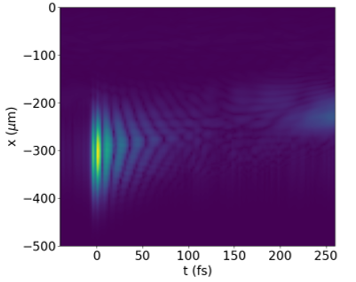}
   \end{tabular}
   \end{center}
   \caption[TDst-Deformed] 
   { \label{fig:TDst-Deformed} 
 Spatio-temporal intensity profiles of the Bragg-diffracted and transmitted beam at \(y = 0\) for the ideal crystal (left) and the deformed crystal (right); color-scale units are arbitrary.
}
\end{figure}

The simulations shown in Fig.~\ref{fig:TDsf-Deformed} and Fig.~\ref{fig:TDst-Deformed} were produced using the following Python scripts, configuration files, and Jupyter notebooks:

\textbf{Left panel in Fig.~\ref{fig:TDsf-Deformed}:}
\begin{itemize}
    \item \texttt{SingleRealization\_C400\_ThermalBump.ipynb}
    \item \texttt{C400\_TD\_9p8keV\_ThermalBump.yaml}
\end{itemize}

\textbf{Central panel in Fig.~\ref{fig:TDsf-Deformed} and left panel in Fig.~\ref{fig:TDst-Deformed}:}
\begin{itemize}
    \item \texttt{run\_parallel\_TD\_Not-deformed.py}
    \item \texttt{C400\_TD\_9p8keV\_No-ThermalBump.yaml}
    \item \texttt{process-parallel-data-omegaC400-No-ThermalBump.ipynb}
\end{itemize}

\textbf{Right panel in Fig.~\ref{fig:TDsf-Deformed} and right panel in Fig.~\ref{fig:TDst-Deformed}:}
\begin{itemize}
    \item \texttt{run\_parallel\_TD\_deformed.py}
    \item \texttt{C400\_TD\_9p8keV\_ThermalBump.yaml}
    \item \texttt{process-parallel-data-omegaC400-ThermalBump.ipynb}
\end{itemize}

Note that during a cross-check of the results published in Ref.~\cite{Krzywinski2022}, 
Section ``3.4 Time-dependent problems: deformed crystals,'' we identified the following mistakes:

\begin{enumerate}
    \item \textbf{Input pulse parameters:}  
    The Gaussian pulse duration should be \textbf{2~fs}, and the RMS transverse size should be 
    \textbf{60\,$\mu$m}, instead of the previously stated \textbf{1~fs} and \textbf{30\,$\mu$m}, respectively.

    \item \textbf{Figure corrections (Fig.~8 in Ref.~\cite{Krzywinski2022}):}  
    The maximum values on the vertical axis should be \textbf{$8\times10^{-5}$}, not \textbf{$8\times10^{-4}$}, 
    and the plots should be \textbf{inverted with respect to the horizontal axis}.
\end{enumerate}

\section{Summary}
 
We have demonstrated that the Fast Fourier Transform Beam Propagation Method (FFT BPM) can be applied to simulate dynamic diffraction effects across a wide range of problems, including scattering from deformed crystals of any shape in Bragg, Laue, or asymmetric geometries. We successfully reproduced results presented in the literature for bent crystals, dislocations and crystals of finite shape simulated using the Takagi–Taupin equations \cite{Samoylova:xn5016, Besedin2014-bk,KowalskiGronkowski,PhysRevB.89.014104}. 

Our FFT BPM method simplifies simulations compared to the TTE method by eliminating the need to solve boundary condition problems. This makes it well-suited for simulating scattering from finite crystals with arbitrary shapes and strain distributions, as well as illumination by X-ray beams with arbitrary angular spectral distributions (e.g., FEL SASE beams). The straightforward algorithm of the FFT BPM method is easy to implement in Python, and the use of FFT ensures fast computation. For instance, simulating 3D reflection from a strained 10~$\mu$m silicon cube on a $300 \times 200 \times 300$ rectangular grid, implemented in a Jupyter notebook and run on a single node of a high-performance computer, takes only a dozen seconds. 

Our method can be naturally implemented on a multicore machine, where single realization reflection objects can be computed in parallel for different parameters, such as photon energy or incidence angle. This makes calculating a series of diffraction patterns by rotating the sample around the angular region near the Bragg peak or performing time-dependent diffraction an embarrassingly parallel problem.

Our compact FFT BPM Python code, with its straightforward implementation, can be easily shared with the X-ray community via GitHub. Here, we present a link to a GitHub repository \cite{fftbpm} containing a Python code package for simulating the examples described in this article. These examples can be modified by adjusting the simulation parameters to address specific dynamical problems of interest in current research. 

\newpage
\appendix
\section{Standardized Simulation Configuration (75 Parameters)}
\label{sec:appendix_params}

The following table provides a one-to-one mapping for the simulation parameters used in our FFT BPM Python package. Following our physical convention, Index 1 is assigned to the top film or surface layer, while Index 2 represents the substrate or bulk material.

\begin{table}[H]
\caption{Complete Parameter Mapping for the 75-Parameter Master Template.}
\label{tab:master_params_75}
\begin{center}       
\begin{tabularx}{\textwidth}{|l|l|X|}
\hline
\textbf{Category} & \textbf{Parameter} & \textbf{Description} \\ \hline
\textbf{Physics \&} & \texttt{omega0, a0} & Photon energy (eV) and Lattice constant (\AA{}). \\
\textbf{Lattice} & \texttt{Miller\_h, k, l} & Reflection indices ($H, K, L$). \\
 & \texttt{delta1, beta1} & Refractive index decrement and absorption (\textbf{Film}). \\
 & \texttt{xrh1, xih1} & Fourier susceptibility components (\textbf{Film}). \\
 & \texttt{delta2, beta2} & Refractive index decrement and absorption (\textbf{Substrate}). \\
 & \texttt{xrh2, xih2} & Fourier susceptibility components (\textbf{Substrate}). \\ \hline
\textbf{Grid \&} & \texttt{xgrid, ygrid} & Number of grid points in $X$ and $Y$. \\
\textbf{Geometry} & \texttt{res\_x, res\_y} & Manual resolution overrides ($0 = \text{auto}$). \\
 & \texttt{xxmax, yymax} & Total simulation window dimensions ($\mu$m). \\
 & \texttt{asymm\_angle} & Asymmetry angle ($0^\circ$ = Bragg, $90^\circ$ = Laue). \\
 & \texttt{thickness} & Substrate thickness ($\mu$m). \\
 & \texttt{d\_film} & Surface film thickness ($\mu$m). \\
 & \texttt{width} & Finite crystal width ($\mu$m). \\
 & \texttt{xs, x00} & Crystal lateral shift and global origin offset ($\mu$m). \\
 & \texttt{geometry} & \texttt{'not\_from\_file', 'from\_file', 'hemi-sphere'}. \\
 & \texttt{geometry\_file} & Path to \texttt{.npy} file for arbitrary shapes. \\ \hline
\textbf{Numerical} & \texttt{M} & Number of integration steps along $Z$. \\
\textbf{Integration} & \texttt{tpad} & FFT padding for wrap-around prevention. \\
 & \texttt{Zstep\_factor} & Step size stability multiplier. \\
 & \texttt{method, mode} & Integration method (\texttt{Euler/Verlet}) and mode (\texttt{TIDP/TDP}). \\
 & \texttt{zsep} & Interval for saving $Z$-layers. \\ \hline
\textbf{Beam} & \texttt{beam, waist} & Beam profile type and FWHM/waist size ($\mu$m). \\
\textbf{Properties} & \texttt{slit\_x, slit\_y} & Simulation slit dimensions ($\mu$m). \\
 & \texttt{time\_dependent} & \texttt{'None'} or \texttt{'Temporal'}. \\
 & \texttt{tgrid, tmax} & Number of time steps and total time range. \\
 & \texttt{sigma\_t} & Pulse duration (fs/ps). \\ \hline
\textbf{Deformation} & \texttt{deformation} & Type (\texttt{Bent Spectrometer, ScrewDislocation, etc.}). \\
\textbf{\& Dislocations}& \texttt{strain, R} & Uniform strain and Bending radius (m). \\
 & \texttt{vC100, vC, vSi} & Poisson ratios for specific materials. \\
 & \texttt{x0d, y0d, z0d} & Dislocation core coordinates ($\mu$m). \\
 & \texttt{phid, burgers\_vector} & Dislocation angle and Burgers vector components. \\
 & \texttt{def\_amplitude} & Periodic deformation amplitude (\AA{}). \\
 & \texttt{def\_period} & Periodic deformation period ($\mu$m). \\ \hline
\textbf{Program} & \texttt{nthread\_fft} & Number of CPU threads for FFT execution. \\
\textbf{Settings} & \texttt{quiet\_mode} & Suppression of console output. \\
 & \texttt{store\_fields} & Flag to save final electric field distributions. \\ \hline
\end{tabularx}
\end{center}
\end{table}
   
\FloatBarrier
  
\acknowledgments 
 
This work was supported by the U.S. Department of Energy (DOE) Contract No. DE-AC02-76SF00515 with SLAC.
\bibliography{report} 
\bibliographystyle{spiebib} 

\end{document}